\documentclass[twoside,twocolumn,9pt]{article}
\usepackage{extsizes}
\usepackage[super,sort&compress,comma]{natbib} 
\usepackage[version=3]{mhchem}
\usepackage[left=1.5cm, right=1.5cm, top=1.785cm, bottom=2.0cm]{geometry}
\usepackage{balance}
\usepackage{mathptmx}
\usepackage{sectsty}
\usepackage{graphicx} 
\usepackage{lastpage}
\usepackage[format=plain,justification=justified,singlelinecheck=false,font={stretch=1.125,small,sf},labelfont=bf,labelsep=space]{caption}
\usepackage{float}
\usepackage{fancyhdr}
\usepackage{fnpos}
\usepackage[english]{babel}
\addto{\captionsenglish}{%
  \renewcommand{\refname}{Notes and references}
}
\usepackage{array}
\usepackage{droidsans}
\usepackage{charter}
\usepackage[T1]{fontenc}
\usepackage[usenames,dvipsnames]{xcolor}
\usepackage{setspace}
\usepackage[compact]{titlesec}
\usepackage{xr-hyper}
\usepackage{hyperref}
\usepackage{cleveref}
\makeatletter
\newcommand*{\addFileDependency}[1]{
  \typeout{(#1)}
  \@addtofilelist{#1}
  \IfFileExists{#1}{}{\typeout{No file #1.}}
}
\makeatother
\newcommand*{\myexternaldocument}[1]{%
    \externaldocument{#1}%
    \addFileDependency{#1.tex}%
    \addFileDependency{#1.aux}%
}

\myexternaldocument{SI}

\usepackage{upgreek}
\usepackage{mathtools}
\usepackage{booktabs}

\usepackage{nomencl}
\makenomenclature
\usepackage{etoolbox}
\renewcommand\nomgroup[1]{%
  \item[\bfseries
  \ifstrequal{#1}{P}{Parameters}{%
  \ifstrequal{#1}{V}{Variables}{%
  \ifstrequal{#1}{I}{Indices}{}}}%
]}
\usepackage{siunitx}
\usepackage{collcell}
\newcolumntype{s}{>{\collectcell\unit}c<{\endcollectcell}}
\newcommand{\nomunit}[1]{%
\renewcommand{\nomentryend}{\hspace*{\fill}\si{#1}}}

\definecolor{cream}{RGB}{222,217,201}
\allowdisplaybreaks
\begin{document}

\newcommand{\vel}{\mathbf{v}}
\newcommand{\mvel}{\vel^{\mathrm{m}}}
\newcommand{\vvel}{\vel^{\mathrm{v}}}
\newcommand{\tvvelwithconvection}{t^{\mathrm{CV}}}
\newcommand{\solventlabel}{\mathrm{s}}
\newcommand{\svel}{\vel^{\solventlabel}}
\newcommand{\Nflux}{N}
\newcommand{\Jflux}{J}
\newcommand{\jlab}{j}
\newcommand{\chempot}{\upmu}
\newcommand{\conc}{c}
\newcommand{\elpot}{\Phi}
\newcommand{\Mflux}{\Nflux^{\mathrm{m}}}
\newcommand{\Vflux}{\Nflux^{\mathrm{v}}}
\newcommand{\Sflux}{\Nflux^{\solventlabel}}
\newcommand{\pmv}{\upnu}
\newcommand{\mobility}{u}
\newcommand{\entropy}{R}
\newcommand{\onsager}{\mathcal{L}}
\newcommand{\charge}{q} 
\newcommand{\density}{\uprho}
\newcommand{\efield}{\boldsymbol{E}}

\pagestyle{fancy}
\thispagestyle{plain}
\fancypagestyle{plain}{
\renewcommand{\headrulewidth}{0pt}
}

\makeFNbottom
\makeatletter
\renewcommand\LARGE{\@setfontsize\LARGE{15pt}{17}}
\renewcommand\Large{\@setfontsize\Large{12pt}{14}}
\renewcommand\large{\@setfontsize\large{10pt}{12}}
\renewcommand\footnotesize{\@setfontsize\footnotesize{7pt}{10}}
\makeatother

\renewcommand{\thefootnote}{\fnsymbol{footnote}}
\renewcommand\footnoterule{\vspace*{1pt}%
\color{cream}\hrule width 3.5in height 0.4pt \color{black}\vspace*{5pt}} 
\setcounter{secnumdepth}{5}

\makeatletter 
\renewcommand\@biblabel[1]{#1}            
\renewcommand\@makefntext[1]%
{\noindent\makebox[0pt][r]{\@thefnmark\,}#1}
\makeatother 
\renewcommand{\figurename}{\small{Fig.}~}
\sectionfont{\sffamily\Large}
\subsectionfont{\normalsize}
\subsubsectionfont{\bf}
\setstretch{1.125} 
\setlength{\skip\footins}{0.8cm}
\setlength{\footnotesep}{0.25cm}
\setlength{\jot}{10pt}
\titlespacing*{\section}{0pt}{4pt}{4pt}
\titlespacing*{\subsection}{0pt}{15pt}{1pt}

\fancyfoot{}
\fancyfoot[LO,RE]{\vspace{-7.1pt}\includegraphics[height=9pt]{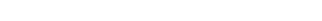}}
\fancyfoot[CO]{\vspace{-7.1pt}\hspace{13.2cm}\includegraphics{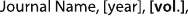}}
\fancyfoot[CE]{\vspace{-7.2pt}\hspace{-14.2cm}\includegraphics{head_foot/RF}}
\fancyfoot[RO]{\footnotesize{\sffamily{1--\pageref{LastPage} ~\textbar  \hspace{2pt}\thepage}}}
\fancyfoot[LE]{\footnotesize{\sffamily{\thepage~\textbar\hspace{3.45cm} 1--\pageref{LastPage}}}}
\fancyhead{}
\renewcommand{\headrulewidth}{0pt} 
\renewcommand{\footrulewidth}{0pt}
\setlength{\arrayrulewidth}{1pt}
\setlength{\columnsep}{6.5mm}
\setlength\bibsep{1pt}

\makeatletter 
\newlength{\figrulesep} 
\setlength{\figrulesep}{0.5\textfloatsep} 

\newcommand{\topfigrule}{\vspace*{-1pt}%
\noindent{\color{cream}\rule[-\figrulesep]{\columnwidth}{1.5pt}} }

\newcommand{\botfigrule}{\vspace*{-2pt}%
\noindent{\color{cream}\rule[\figrulesep]{\columnwidth}{1.5pt}} }

\newcommand{\dblfigrule}{\vspace*{-1pt}%
\noindent{\color{cream}\rule[-\figrulesep]{\textwidth}{1.5pt}} }

\makeatother

\twocolumn[
  \begin{@twocolumnfalse}
{\includegraphics[height=30pt]{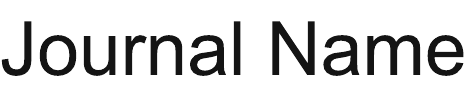}\hfill\raisebox{0pt}[0pt][0pt]{\includegraphics[height=55pt]{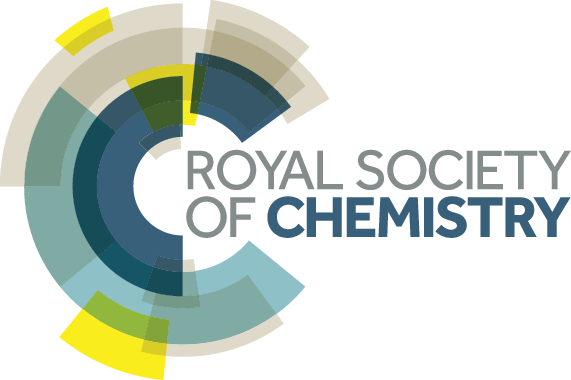}}\\[1ex]
\includegraphics[width=18.5cm]{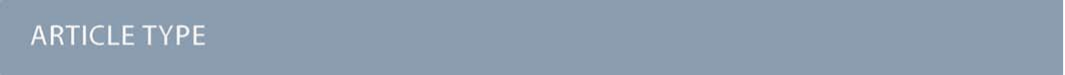}}\par
\vspace{1em}
\sffamily
\begin{tabular}{m{4.5cm} p{13.5cm} }

\includegraphics{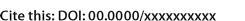} & \noindent\LARGE{\textbf{A Volume-based Description of Transport in Incompressible Liquid Electrolytes and its Application to Ionic Liquids}$^\dag$} \\
\vspace{0.3cm} & \vspace{0.3cm} \\

 & \noindent\large{Franziska Kilchert,\textit{$^{ab}$} Martin Lorenz,\textit{$^{c}$} Max Schammer,\textit{$^{ab}$} Pinchas N\"urnberg,\textit{$^{c}$} Monika Sch\"onhoff,\textit{$^{c}$} Arnulf Latz\textit{$^{abd}$} and Birger Horstmann$^{\ast}$\textit{$^{abd}$}} \\

\includegraphics{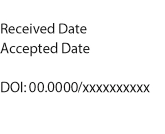} & \noindent\normalsize{Transference numbers play an important role for understanding the dynamics of electrolytes and assessing their performance in batteries. Unfortunately, these transport parameters are difficult to measure in highly concentrated, liquid electrolytes such as ionic liquids. Also, the interpretation of their sign and magnitude has provoked an ongoing debate in the literature further complicated by the use of different language. In this work, we highlight the role of the reference frame for the interpretation of transport parameters using our novel thermodynamically consistent theory for highly correlated electrolytes. We argue that local volume conservation is a key principle in incompressible liquid electrolytes and use the volume-based drift velocity as reference. We apply our general framework to electrophoretic NMR experiments. For ionic liquid based electrolytes, we find that the results of the eNMR measurements can be best described using this volume-based description. This highlights the limitations of the widely used center-of-mass reference frame which for example forms the basis for molecular dynamics simulations - a standard tool for the theoretical calculation of transport parameters. It shows that the assumption of local momentum conservation is incorrect in those systems on the macroscopic scale.} \\

\end{tabular}

 \end{@twocolumnfalse} \vspace{0.6cm}

  ]

\renewcommand*\rmdefault{bch}\normalfont\upshape
\rmfamily
\section*{}
\vspace{-1cm}


\footnotetext{\textit{$^{a}$~German Aerospace Center, Wilhelm-Runge-Stra{\ss}e 10, 89081 Ulm, Germany. E-mail: birger.horstmann@dlr.de}}
\footnotetext{\textit{$^{b}$~Helmholtz Institute Ulm, Helmholtzstra{\ss}e 11, 89081 Ulm, Germany.}}
\footnotetext{\textit{$^{c}$~University of M\"unster, Corrensstra{\ss}e 28/30, 48149 M\"unster, Germany.}}
\footnotetext{\textit{$^{d}$~Ulm University, Albert-Einstein-Allee 47, 89081 Ulm, Germany.}}

\footnotetext{\dag~Electronic Supplementary Information (ESI) available:
	\\
	See DOI: 00.0000/00000000.}




\section{Introduction}

The large majority of technical applications for energy storage employ concentrated liquid electrolytes, including standard electrolytes with around 1 mol/l of solute and such with higher salt concentrations.\cite{Zheng2017} With modeling we can provide deeper insights into the internal transport processes. Here, transport parameters like the transference number play a major role as performance indicator. In principle, the transference number $t_\alpha=I_\alpha/I$ of a species is defined as the fraction of current $I_\alpha$, carried by the respective species, with respect to the overall current $I$. However, the determination of transference numbers in highly concentrated electrolytes such as ionic liquids (ILs) is challenging.

A multitude of measurement techniques have been developed over time, in part assisted by modeling approaches. This variety together with diverse nomenclature hampers the comparability of experimental results. One prominent example being the recent discovery of negative transference numbers for several highly concentrated electrolytes and the resulting debate.\cite{Ma1995,Ferry1998,Pesko2017,Brinkkoetter2018,Gouverneur2018,Shah2019,Molinari2019_1,Molinari2019_2} Next, we summarize the diversity of experimental and theoretical methods in this domain.

On the experimental side, electrochemical methods are the most widespread. Here, potentiostatic polarization (PP) developed by Bruce and Vincent and based on the classical Hittorf method\cite{Kaimakov1966,Bruce1987_1,Bruce1987_2,Bruce1992} and galvanostatic polarization (GP) from Newman and co-workers\cite{Ma1995} are prominent techniques besides some more recent developments that aim at reducing the number of experiments needed to obtain a full set of independent transport parameters.\cite{Landesfeind2016,Ehrl_2017_1,Ehrl_2017_2,Landesfeind2019} Another approach uses very-low frequency impedance spectroscopy (VLF-EIS).\cite{Wohde2016,Vargas-Barbosa2019,Saelzer2020,Ho2022} It is common sense that ideal ion dissolution is no valid assumption in concentrated electrolytes and, thus, quantities obtained from PFG-NMR via the Nernst-Einstein relation cannot be considered "true" transference numbers. However, they are still calculated in many cases as comparative figure.\cite{Zugmann2011,Wohde2016,Pesko2017,Saelzer2020} The relatively new technique of electrophoretic NMR (eNMR), however, is a valuable tool for analyzing highly concentrated electrolytes.\cite{Gouverneur2015,Gouverneur2018,Rosenwinkel2019} Here, the mobilities of each species in the system can be measured directly and, thus, be used to determine transference numbers.

Physiochemical models, on the other hand, can supplement experimental techniques and are a valuable tool to rationalize the different definitions for transference numbers proposed in the literature. Different parameter definitions and experimental set-ups can be exactly quantified with sufficient effort, \emph{e.g.} the relation between VLF-EIS and eNMR.\cite{Pfeifer2021} 

Transference numbers depend on the adopted frame of reference determined by the underlying theory. The choice of the reference frame specifies the ion flux densities which are defined relative to the corresponding drift velocity of the reference frame. This well known fact has been described extensively in the literature.\cite{DeGroot1962,Tyrrell1984,Haase1990} Kirkwood et al. first stated coherently the reference frame dependence of equations to test Onsager's reciprocal relations for isothermal diffusion in multi-component systems.\cite{Kirkwood1960} Flux densities and constraints are given in the local center-of-mass (CM), local volume and local species-based reference frame. However, no electric currents were considered. Miller showed in great detail the connection between Onsager coefficients and transport parameters for binary to N-component systems focusing on the species-based reference frame.\cite{Miller1966a,Miller1967b,Miller1967c} Transformations between reference frames are also considered early on, \emph{e.g.} by Schönert who discusses Onsager coefficients in the solvent- and CM-based frames.\cite{Schoenert1984} In the battery-modelling community, the most widely used descriptions are based on the center-of-mass motion,\cite{Latz2011,Newman1965,newman2012electrochemical} or the solvent-based species-frame.\cite{Newman1967,newman2012electrochemical,Mistry_2022} In contrast, the volume-based description has been discussed only rarely in the literature.\cite{Newman1973,Brenner2005,Brenner2003,Goyal_2017,Liu2014,Wang2021}

For a certain application one reference frame is more convenient for describing observed phenomena than another. For example, in the presence of an excess solvent species, \emph{e.g.} in aqueous electrolytes or polymer electrolytes, the species velocity of the solvent is the natural reference. Also, the aforementioned Hittorf method for determining transference numbers relates to the solvent-fixed reference frame\cite{Miller1966a} and models based on the Stefan-Maxwell theory, like the widespread concentrated solution theory are formulated in this frame.\cite{Doyle1994,Nyman2008,Hou2020} Another choice for the reference frame often used in the literature is the mass-based frame where the ion flux densities are expressed relative to the center-of-mass (CM) motion. Because the corresponding drift velocity equals the bulk momentum per unit mass, this description is inherently related to momentum balance and applies to highly viscous liquids where momentum dissipation due to friction forces must be taken into account.\cite{doi:10.1080/00268978400102441} For describing bulk transport in IL electrolytes using an Onsager approach, the mass-based description is the natural choice of reference.\cite{schammer2020theory} This frame also forms the reference description used in molecular dynamics (MD) simulations.\cite{Shao2022} Here, structural and dynamical properties can be directly accessed from the atomic or molecular structure of the system evolving with time. Then, parameters like the transference number can be calculated using analytical tools.\cite{Wheeler2004_1,Wheeler2004_2,Molinari2019_1}

With our work we want to shed some light on recent discussions about reference frames and corresponding transference numbers in concentrated, liquid electrolytes to help improving the mutual understanding. Here, we use thermodynamically consistent modeling based on an Onsager approach which assumes that the thermodynamic flux densities are linear functions of the electrochemical forces.\cite{DeGroot1962,schammer2020theory} Thermodynamic flux densities are usually defined relative to the internal dynamics of some excess bulk quantity. These internal flux densities differ from the external flux densities by a convective correction, where the choice of the drift velocity defines the corresponding frame of reference. In this flux-explicit description, the Onsager coefficients determine the transport parameters. As consequence, all transport parameters are defined with respect to a specific frame of reference, \emph{i.e.} choice for the drift velocity. In the beginning of our theory section, we recapitulate the basic frame-dependent flux density definitions. As we show in detail, using the volume-based drift velocity as reference can be a convenient ansatz for electrolytes where the volume does not change under pressure variation, i.e. incompressible electrolytes. Since electrolytes are hardly compressible this is usually a good approximation.\cite{Dreyer2013} In contrast to the CM-based frame of reference, this description models species transport via volume-preserving flux densities and, thus, facilitates the transport equation for the drift velocity. Furthermore, we derive simple transformation rules between the descriptions.

In highly concentrated electrolytes, strong ion correlations and conservation laws such as charge conservation lead to constraints on the dynamics of the electrolyte species. In the volume-based description volume conservation replaces the typical momentum conservation from CM-based frames as governing principle. The resulting constraints transfer to the thermodynamic flux densities such that only a reduced number of internal flux densities is independent (\mbox{N-1} in case of N species). Furthermore, this implies that not all Onsager coefficients are independent. Thus, only a reduced number of transport parameters is independent and can be measured independently. For example, in the case of transference numbers there are two constraints. First, one ion flux density determines a designated species. Second, the transference numbers of the remaining species sum up to unity.\cite{schammer2020theory} As consequence, there exist only \mbox{N-2} independent transference numbers in the internal description of an N-component electrolyte solution. This can lead to confusion regarding the sign and magnitude of the transport parameters when different descriptions from the literature are compared. Hence, it is important to clearly state the reference frame for the transport description. Interestingly, as pointed out by Harris,\cite{harris2018comment} these issues were clearly understood for molten salt systems already in the 1950s.\cite{Schoenert1962,Sinistri1962,klemm1964transport,klemm1960phanomenologie,haase1973deutung,Hussey01081987,sundheimfused,sundheim1956transference,KAWAMURA19711151,Duke01051957}

In this work, we rationalize the relevance of reference frames on the sign and magnitude of transference numbers, identify the set of independent parameters, and discuss alternative definitions for the transference numbers. Our joint theoretical / experimental work described here focuses on theoretical aspects and is accompanied by a second publication of the same authors which focuses on the experimental aspects.\cite{Lorenz2022} Here, we show that our volume-based description is for example applicable with porous electrode theory as electrolyte modeling framework. For incompressible electrolytes the volume-based representation facilitates the volume velocity equation. We, finally, apply it to eNMR experiments where the volume-based reference frame turns out to be especially beneficial due to the vanishing volume flux boundary condition of the system. Furthermore, we show that incompressible electrolytes are better described with local volume conservation and how we can explain experimental findings with our theory. We build the theoretical framework based on irreversible thermodynamics and focus on transference numbers. Our complementary publication, see Ref.~\citenum{Lorenz2022}, focuses on the experimental aspects of our collaboration and provides eNMR and density data to experimentally validate the theoretical concepts.

\section{Theory}
\label{sec:main_theory}

Recently, Schammer et al. proposed a novel transport theory for highly correlated electrolytes.\cite{schammer2020theory} It captures the cell dynamics of electrochemical devices on a macro-scale, \emph{e.g.} in the electroneutral bulk,\cite{schammer2020theory} but also for strong electrostatic correlations in crowded environments, \emph{e.g.} in the electrochemical double layer (EDL).\cite{schammer2021role,Hoffmann2018} The focal quantity in this modeling framework is the Helmholtz free energy, which incorporates the material-specific properties, and determines the description of the system in the form of constitutive equations. Here, we apply the framework outlined in great detail in Ref.~\citenum{schammer2020theory} and focus on the transport parameters.

We structure this theory part as follows. First, in \cref{sec:intro_theory}, we briefly summarize how the flux densities are defined in different internal reference frames. We put emphasis on the dominant conservation laws and resulting constraints appearing in strongly correlated electrolytes, which reduce the number of independent parameters, and on the frame-dependence of the transport parameters. This is followed by two sections in which we focus on two different reference frames. In \cref{sec:mass_description}, we outline the canonical approach where the center-of-mass motion is used as drift velocity. In \cref{sec:VolBasedDescription}, we discuss an alternative description based on the volume motion. In \cref{sec:frame_transformation}, we show how to transform between reference frames. Finally, in \cref{sec:transference_number}, we focus on transference numbers and their reference frame dependence.

To facilitate the readability of this theory section, we provide a list of symbols at the end of the manuscript.

\subsection{Internal Frames of Reference}
\label{sec:intro_theory}

Our transport theory describes the evolution of state variables. It is based on the dynamics of species flux densities that relate to some reference frame. Mutual couplings between these flux densities, \emph{e.g.} due to balancing laws, impose flux constraints and reduce the number of independent fluxes.

Our framework applies to electrolyte mixtures composed of N different constituent species. We express the corresponding flux densities as $\Nflux_\alpha^{\uppsi} = \conc_\alpha (\vel_\alpha-\vel^{\uppsi})$, where $\conc_\alpha$ are the species amount concentrations, via the relative motion between the species velocities $\vel_\alpha$ and the bulk motion of some internal electrolyte quantity, \emph{i.e.} drift velocity, $\vel^{\uppsi}$. Usually, the drift velocity $\vel^{\uppsi}$ is defined as linear function of some internal electrolyte quantity $\uppsi_\alpha$ which exists in bulk-excess, $\vel^{\uppsi}=\sum_{\alpha=1}^{\ce{N}} \uppsi_\alpha \cdot \vel_{\alpha}$. Thus, in our description, $\vel^{\uppsi}$ constitutes an “internal” reference frame, which highlights the special role of the drift velocity. Here, we assume that the expansion coefficients are normalized,  $\sum_{\alpha=1}^{\ce{N}} \uppsi_\alpha = 1$. By construction, this implies a universal flux constraint,\cite{Goyal_2017}  
\begin{equation}
	\label{eq:universal_flux_constraint}
	\sum_{\alpha=1}^{\ce{N}} \frac{\uppsi_\alpha}{ \conc_\alpha} \cdot \Nflux_\alpha^{\uppsi} = 0.
\end{equation}
As consequence, in an electrolyte mixture composed of N different species, there exist only N-1 independent flux densities $\Nflux_\alpha^{\uppsi}$.

In the literature, many different expressions for the drift velocity exist.\cite{Kirkwood1960,DeGroot1962,Tyrrell1984} The relevance of specific choices depends upon the physical system, \emph{i.e.} the experimental set-up and boundary conditions. Common examples include the velocity of the center-of-mass motion,\cite{Latz2011,schammer2020theory,Newman1965,newman2012electrochemical} the volume-based velocity,\cite{Newman1973,Brenner2005,Brenner2003,Goyal_2017,Liu2014,Wang2021} or internal species velocities like the solvent motion.\cite{Newman1967,newman2012electrochemical,Mistry_2022} However, accounting for bulk motion of the electrolyte in the dynamical description implies that $\vel^{\uppsi}$ becomes a dynamical variable. This requires that the system of transport equations must be supplemented by an additional drift velocity equation.

The widely used CM description is based on the CM drift velocity $\mvel = \sum_{\alpha=1}^{\ce{N}}\density_\alpha \cdot \vel_\alpha/\density$, where 
\begin{equation}
\label{eq:dens}
    \density = \sum_{\alpha=1}^{\ce{N}} \density_\alpha = \sum_{\alpha=1}^{\ce{N}} M_\alpha \cdot \conc_\alpha
\end{equation} 
is the total mass density with $M_\alpha$ the molar masses (here, $\uppsi_\alpha=\density_\alpha/\density$). Note that the drift velocity $\mvel$ is equivalent to the momentum per unit mass. Hence, the CM frame naturally relates to momentum balance and, thus, has some conceptual advantages. The corresponding species flux densities $\Mflux_\alpha$ read
\begin{gather} 
	\label{eq:massflux}
	\Mflux_\alpha = \conc_\alpha \left( \vel_\alpha - \mvel \right).
	\shortintertext{In the CM frame, the universal flux constraint from above, see \cref{eq:universal_flux_constraint} becomes }
	\label{eq:massfluxconstr}
	\sum_{\alpha=1}^{\mathrm{N}} M_{\alpha}\cdot \Mflux_\alpha = 0.
\end{gather}

An alternative frame of reference is the volume-based description, which naturally relates to volume conservation. This internal frame is based on the volume fractions of the electrolyte species, \emph{i.e.} $\uppsi_\alpha =  \conc_\alpha\pmv_\alpha$. Here, the quantity $\pmv_\alpha$ is the partial molar volume of species $\alpha$ and $c_\alpha\pmv$ are the volume fractions. Because the volume is a homogeneous function of the particle number we can use Euler's theorem for homogeneous functions,\cite{Reiss1996} from which follows\cite{schammer2020theory}
\begin{equation} 
	\label{eq:volconstr}
	\sum_{\alpha=1}^{\mathrm{N}} \conc_\alpha \cdot \pmv_\alpha = 1.
\end{equation}
This "Euler equation for the volume" expresses the volume-filling property of the multi-component electrolyte.  The corresponding volume drift velocity in this reference frame is $\vvel = \sum_{\alpha=1}^{\ce{N}} \pmv_\alpha\conc_\alpha \cdot \vel_\alpha$, and gives rise to flux densities
\begin{gather} 
	\label{eq:volflux}
	\Vflux_\alpha = \conc_\alpha \left( \vel_\alpha - \vvel \right).
	\shortintertext{The universal flux constraint, see \cref{eq:universal_flux_constraint}, in this frame reads}
	\label{eq:volfluxconstr}
	\sum_{\alpha=1}^{\mathrm{N}} \pmv_\alpha \cdot \Vflux_\alpha = 0.
\end{gather}
The Euler equation for the volume, \cref{eq:volconstr}, constitutes a kinematic relation between the volume phases of the electrolyte, which plays a fundamental role for concentrated electrolytes. This is true both for the bulk, where it leads to a description for the spatial variation of the drift velocity,\cite{schammer2020theory} and for the EDL, where it stabilizes the bulk structure against unbalanced Coulomb-attraction of the ions.\cite{schammer2021role}

Finally, we discuss the description based on an internal species velocity, \emph{i.e.} where the drift velocity is given by the velocity of one designated electrolyte species. By convention, we set $\alpha=1$ for the  designated species  such that $\svel=\vel_1$ (thus, $\uppsi_\alpha = \delta_{\alpha}^{1}$). This description is a common choice in the presence of a dominant solvent species, where $\svel$ defines the "solvent-fixed frame", or for polymer electrolytes, where the polymer would be the designated species. The flux densities then take the form
\begin{equation}
    \label{eq:speciesflux}
    \Sflux_\alpha = \conc_\alpha \left( \vel_\alpha - \svel \right) = \conc_\alpha \left( \vel_\alpha - \vel_1 \right),
\end{equation}
and the universal flux constraint reduces to $\Sflux_1 = 0$, see \cref{eq:universal_flux_constraint}.

\subsection{Mass-based Description}
\label{sec:mass_description}

In this section, we focus on the mass-based description, \emph{i.e.} the reference frame constituted by the center of mass motion (see \cref{sec:intro_theory}). 

The framework of rational thermodynamics obeys the second law of thermodynamics using the entropy production rate $\entropy$, where thermodynamic consistency demands that the quantity $\entropy$ is strictly non-negative. Hence, the entropy production rate measures the deviation from thermodynamic equilibrium, and expresses balance of entropy production. As we show in Ref.~\cite{schammer2020theory}, $\entropy$ comprises a contribution due to internal friction, \emph{i.e.} viscosity, and a flux term 
\begin{equation} 
\label{eq:entropy_full}
\entropy^{\ce{flux}} 
 = - \Jflux^{\ce{m}} \nabla\elpot- \sum_{\alpha=1}^{\mathrm{N}} \Mflux_\alpha
 \cdot\nabla \chempot_\alpha
  = - \sum_{\alpha=1}^{\mathrm{N}} \Mflux_\alpha \cdot \nabla \chempot_\alpha^{\ce{el}}.
\end{equation}
Here, $\chempot_\alpha$ are the specific chemical potentials, and $\chempot_\alpha^{\mathrm{el}}=\chempot_\alpha+Fz_\alpha\elpot$ are the electrochemical potentials ($F$ denotes the Faraday constant and $\elpot$ the electric potential).  The chemical potentials follow from our model for the free energy $\upvarphi_{\ce{H}}$ presented in Ref.~\citenum{schammer2020theory} via $\chempot_\alpha = \partial(\density\upvarphi_{\ce{H}})/\partial\conc_\alpha$. Above, we used charge continuity and expressed the electric current density via $\Jflux^{\ce{m}}=\sum_{\alpha=1}^{\ce{N}}Fz_\alpha\Nflux^{\ce{m}}_\alpha$. Next, we evaluate the flux constraint \cref{eq:massflux} and reduce the flux-explicit expansion of $\entropy$ appearing on the right of \cref{eq:entropy_full} to the set of independent fluxes,
\begin{gather}
\label{eq:entropy}
\entropy^{\ce{flux}} 
 = - \sum_{\alpha=2}^{\mathrm{N}} \Mflux_\alpha \cdot \nabla \tilde{\chempot}_\alpha^{\mathrm{m,el}},
 \intertext{where we introduced}
 \label{eq:reduction_mu_z}
\tilde{\chempot}_\beta^\mathrm{m,el} {=} \chempot_\beta {-} M_\alpha/ M_1{\cdot} \chempot_1 {-} F\tilde{z}_\alpha^\mathrm{m}\elpot,
\ \  \textnormal{and} \ \ 
\tilde{z}_\alpha^\mathrm{m} {=} z_\alpha {-} M_\alpha / M_1 {\cdot} z_1.
\end{gather}

The thermodynamic  requirement that $\entropy$ is non-negative determines the constitutive modeling of the thermodynamic variables and flux densities. Here, we ensure non-negativity of \cref{eq:entropy} using an Onsager approach for the flux densities,
\begin{equation}
    \label{eq:onsager}
	\Mflux_\alpha = - \sum_{\beta=2}^\mathrm{N} \onsager_{\alpha \beta}^{\ce{m}} \cdot \nabla \tilde{\chempot}_\beta^\mathrm{m},
\end{equation}
where the corresponding symmetric Onsager matrix $\onsager_{\alpha \beta}^{\ce{m}}$ is semi-positive definite (see \cref*{eq:SI_onsager} in the ESI$^{\dag}$ for a matrix expression).

All macroscopic transport parameters follow directly from the Onsager matrix. Note that the sum on the right side starts at counter two, which restricts the number of independent Onsager coefficients, \emph{i.e.} transport parameters (the choice for the electrolyte species which is set to $\alpha=1$ can be motivated with respect to the physical set-up, see Ref.~\citenum{schammer2020theory}).  Altogether, symmetry and thermodynamic consistency reduce the number of independent transport coefficients appearing in an N-component electrolyte mixture to  N(N+1)/2. This statement is generally true in any internal reference frames. 

The Euler equation for the volume, \cref{eq:volconstr}, and charge continuity $\charge=\sum_{\alpha=1}^{\ce{N}}Fz_\alpha\conc_\alpha$ reduce the set of variables which is necessary for the complete dynamical description of highly concentrated electrolytes.\cite{schammer2020theory} Hence, $\conc_1$ and $\conc_2$ are functions of $(\pmv_\alpha,\charge, \conc_3,\ldots,\conc_{\ce{N}})$ (see \cref*{eq:SI_c1,eq:SI_c2}).$^{\dag}$ Similarly, from the universal flux constraint follows that $\Jflux^{\ce{m}}=\sum_{\alpha=2}^{\ce{N}}F\tilde{z}_\alpha^{\ce{m}}\Nflux_\alpha^{\ce{m}}$, such that $\Nflux_1^{\ce{m}}$ and $\Nflux_2^{\ce{m}}$ are functions of the independent fluxes $(\Jflux^{\ce{m}}, \Nflux_3^{\ce{m}},\ldots,\Nflux_{\ce{N}}^{\ce{m}})$, where 
\begin{gather}
	\label{eq:Jflux}
	\Jflux^{\ce{m}}  = - \kappa^\mathrm{m} \nabla \varphi^\mathrm{m} - \frac{\kappa^\mathrm{m}}{F} \sum_{\beta=3}^\mathrm{N} \frac{t_\beta^\mathrm{m,red}}{\tilde{z}_\beta^\mathrm{m}} \nabla \tilde{\tilde{\chempot}}_\beta^\mathrm{m}
	\shortintertext{and}
	\label{eq:Mflux_explicit}
	\Mflux_\alpha = \frac{t_\alpha^\mathrm{m,red}}{F \tilde{z}_\alpha^\mathrm{m}} \Jflux^\mathrm{m} - \sum_{\beta=3}^\mathrm{N} D_{\alpha\beta}^\mathrm{m} \nabla \tilde{\tilde{\chempot}}_\beta^\mathrm{m}, \qquad \alpha \geq 3.
\end{gather}
Here, $\varphi^\mathrm{m}= \elpot + \tilde{\chempot}_2^\mathrm{m}/F\tilde{z}_2$ is an  electrochemical potential,\cite{latz2015multiscale,newman2012electrochemical} and $\kappa^\mathrm{m}$ and $D_{\alpha\beta}^\mathrm{m}$ are the conductivity and diffusion coefficient (derived from Onsager coefficients, see Ref.~\citenum{schammer2020theory}) in the mass-based frame,\cite{latz2015multiscale} respectively. The transference numbers appearing in \cref{eq:Jflux,eq:Mflux_explicit} satisfy the normalization $\sum_{\alpha=2}^{\ce{N}}t^{\mathrm{m,red}}_\alpha =1$ and constitute the set of N-2 independent parameters in the CM description. Furthermore, we introduced 
\begin{equation}
\tilde{\tilde{\chempot}}_\alpha^\mathrm{m,el} = \tilde{\chempot}_\alpha^\mathrm{m,el} - \tilde{\chempot}_2^\mathrm{m,el} \cdot \tilde{z}_\alpha^\mathrm{m} / \tilde{z}_2^\mathrm{m}
\end{equation}

The resulting set of closed isothermal transport equations reads
\begin{gather}
	\label{eq:transporteq_charge}
	\partial_t  \charge = - \nabla \Jflux^{\mathrm{m}} - \nabla \left( \density \mvel \right),
	\\
	\label{eq:transporteq_mass}
	\partial_t \conc_\alpha = - \nabla \Mflux_\alpha  - \nabla \left( \conc_\alpha \mvel \right), \qquad \alpha \geq 3,
	\\
	\label{eq:CMConv}
	\nabla \mvel 
    = - \sum_{\alpha=1}^{\ce{N}}\pmv_\alpha\cdot \Nflux_\alpha
 = - \frac{\tilde{\pmv}_2^\mathrm{m}}{F \tilde{z}_2^\mathrm{m}} \nabla \Jflux^\mathrm{m} - \sum_{\alpha=3}^\mathrm{N} \tilde{\tilde{\pmv}}_\alpha^\mathrm{m} \cdot \nabla \Mflux_\alpha,
	\\
	\label{eq:poisson}
	\charge = - \upvarepsilon_{\ce{R}}\upvarepsilon_0\Delta\elpot,
\end{gather}
where $\upvarepsilon_{\ce{R}}$ is the relative permittivity and $\upvarepsilon_0$ the vacuum permittivity.  As it was shown in Ref.~\citenum{schammer2020theory}, the convection equation (see \cref{eq:CMConv}) can be derived from the Euler equation for the volumes, see \cref{eq:volconstr}, and can be reduced to an expansion via the independent thermodynamic fluxes $(\Jflux^{\ce{m}}, \Nflux_3^{\ce{m}},\ldots,\Nflux_{\ce{N}}^{\ce{m}})$ using  parameters 
\begin{equation} \label{eq:nuRed}
	\tilde{\pmv}_\alpha^\mathrm{m} = \pmv_\alpha - \pmv_1 \cdot M_\alpha / M_1, \quad \textnormal{and} \quad
  \tilde{\tilde{\pmv}}_\alpha^\mathrm{m} = \tilde{\pmv}_\alpha^\mathrm{m} - \tilde{\pmv}_2^\mathrm{m} \cdot \tilde{z}_\alpha^\mathrm{m} / \tilde{z}_2^\mathrm{m}.
\end{equation}

We emphasize that corrections similar to the relative mass ratios $M_\alpha/M_1$ (as appearing in the quantities $\tilde{\chempot}_\alpha$, $\tilde{z}_\alpha$, ...) emerge naturally in all other descriptions based on different internal frames of references as artefacts of redundant fluxes.

\subsection{Volume-based Description} 
\label{sec:VolBasedDescription}

In this section, we discuss the volume-based frame defined in \cref{sec:intro_theory} (see \cref{eq:volflux,eq:volfluxconstr}). 

Our rationale here is similar to the CM-based description discussed in \cref{sec:mass_description}. The corresponding transport equations are analogous to \cref{eq:transporteq_charge,eq:transporteq_mass,eq:CMConv,eq:poisson}, with the exception of different flux densities and a different drift velocity. Similar to the CM-based description, we apply charge continuity, $\charge = F \sum_{\alpha=1}^\mathrm{N} z_\alpha \conc_\alpha$ and the volumetric flux constraint, \cref{eq:volfluxconstr},  to reduce our description for these flux densities. For this purpose, we introduce reduced quantities involving corrections due to volume ratios,
\begin{equation} \label{eq:volfznutilde}
	\tilde{z}_\alpha^\mathrm{v} = z_\alpha - z_1 \cdot 
	\pmv_\alpha / \pmv_1, \quad \textnormal{and} \quad
	\tilde{\chempot}_\alpha^\mathrm{v} 
	= \chempot_\alpha - \chempot_1 \cdot \pmv_\alpha/\pmv_1.
\end{equation}
Correspondingly, we expand the electric current density via the flux densities, $ \Jflux^\mathrm{v} = F \sum_{\alpha=2}^\mathrm{N} \tilde{z}_\alpha^\mathrm{v} \Vflux_\alpha$, such that 
\begin{gather}
\label{eq:volelcurrreduced}
	\Jflux^\mathrm{v} = - \kappa^\mathrm{v} \nabla \varphi^\mathrm{v} - \frac{\kappa^\mathrm{v}}{F} \sum_{\beta=3}^{\mathrm{N}} \frac{t_\beta^\mathrm{v,red}}{\tilde{z}_\beta^\mathrm{v}} \nabla \tilde{\tilde{\chempot}}_\beta^\mathrm{v},,
 \intertext{with the electric conductivity $\kappa^\mathrm{v}$ and the electrochemical potential $\varphi^\mathrm{v;el}= \elpot + \tilde{\chempot}_2^\mathrm{v}/F\tilde{z}_2$ both in the volume description, and }
	\label{eq:volfluxreduced}
	\Vflux_\alpha = \frac{t_\alpha^\mathrm{v,red}}{F \tilde{z}_\alpha^\mathrm{v}} \Jflux^\mathrm{v} - \sum_{\beta=3}^\mathrm{N} D_{\alpha\beta}^\mathrm{v} \nabla \tilde{\tilde{\chempot}}_\beta^\mathrm{v}, \qquad \alpha \geq 3,
 \intertext{where}
	\label{eq:mutilde2}
	\tilde{\tilde{\chempot}}_\alpha^\mathrm{v} = \tilde{\chempot}_\alpha^\mathrm{v} - \tilde{\chempot}_2^\mathrm{v} \cdot  \tilde{z}_\alpha^\mathrm{v}/\tilde{z}_2^\mathrm{v}. 
\end{gather}
Here, $t_\alpha^\mathrm{v,red}$ denote N-1 transference numbers, where only N-2 transference numbers are independent. $D_{\alpha\beta}^\mathrm{v}$ are the diffusion coefficients in the volume-based frame.

The drift velocity equation in the volume-based description becomes (see \cref*{sec:SI_convection_equation})$^{\dag}$ 
\begin{equation}
	\label{eq:special_volvel_zero}
	\nabla \vvel 
	= - \sum_{\alpha=1}^{\mathrm{N}} \pmv_\alpha \cdot \nabla\Vflux_\alpha
	= \sum_{\alpha=1}^{\mathrm{N}} \Vflux_\alpha \cdot \nabla \pmv_\alpha
	= \left( \sum_{\alpha=1}^{\mathrm{N}} \Vflux_\alpha \cdot \frac{\partial \pmv_\alpha}{\partial p} \right) \nabla p ,
\end{equation}
where we used the flux relation imposed by \cref{eq:volfluxconstr} and $p$ is the pressure. In the ESI,$^{\dag}$ we derive a reduced description for the drift velocity equation in the volume-based description (see \cref*{eq:SI_volumeconv_reduced}).

When the partial molar volumes do not depend on pressure, \emph{i.e.} for incompressible electrolytes, the volume-based drift velocity is spatially constant,\cite{DeGroot1962}
\begin{equation} \label{eq:conveqzero}
	\nabla \vvel = 0.
\end{equation}
This relation between incompressibility and constant drift velocity $\vel^{\uppsi}$ is generally true in mono-component liquids. In contrast, for multi-component mixtures, it is only true in the volume-based description. Nevertheless, in the literature, $\nabla \vel^{\uppsi}=0$ is often used to define incompressible electrolytes, regardless of the specific frame used. This can be a bad  approximation for highly concentrated electrolytes.\cite{schammer2020theory} However, when \cref{eq:conveqzero} holds, then the drift velocity is completely determined by the boundary conditions. In the case of reactions occurring at the electrodes, then $\nabla\vvel=\sum_{\alpha=1}^{\ce{N}}\pmv_\alpha r_\alpha$, where $r_\alpha$ is the reaction rate of the species $\alpha$ at the electrodes.

\subsection{Flux Transformations Between Different Reference Frames in Incompressible Electrolytes}
\label{sec:frame_transformation}

In this section, we state transformation rules for the drift velocities, flux densities and current densities between different reference frames. To improve readability we state the results in the main text and provide a complete derivation in the ESI, see \cref*{sec:SI_DerivTransformRules_vm}.$^{\dag}$ Here, we focus on the two cases of the CM-based description and the volume-based description discussed in \cref{sec:mass_description,sec:VolBasedDescription}. In addition, we discuss transformation rules with respect to the species-based reference frame in the ESI, see \cref*{sec:SI_DerivTransformRules_vs}. From now on, we focus on incompressible electrolytes where the partial molar volumes do not depend on the pressure and assume that $\nabla\pmv_\alpha=0$.

It is a fundamental assumption of physics that the dynamical evolution of a macroscopic system is independent from the internal description, \emph{i.e.} the choice of reference frame. This implies that frame transformations are symmetries of the macroscopic description. However, in contrast to the macroscopic description, internal quantities, \emph{e.g.} flux densities, depend on the reference frame, and simple frame transformations exist.

The transformation of the drift velocities between the the CM-based description and  the volume-based description reads (see \cref*{sec:SI_DerivTransformRules_vm} in the ESI)
\begin{gather}
    \label{eq:frame_trf}
	\vvel - \mvel 
    = \frac{M_1}{\density\pmv_1}\sum_{\beta=2}^\mathrm{N} \tilde{\pmv}_\beta^{\ce{m}} \cdot \Vflux_\beta 
    = \sum_{\beta=2}^\mathrm{N} \tilde{\pmv}_\beta^{\ce{m}} \cdot \Mflux_\beta.
\end{gather}
This transformation rule for the drift velocities determines the transformation rule for the flux densities $\Nflux^\uppsi_\alpha$ and electric current densities $\Jflux^{\uppsi}$,
\begin{gather}
	\label{eq:Jflux_trf}
    \Jflux^\mathrm{m} - \Jflux^\mathrm{v}
    = \charge \frac{M_1}{\density\pmv_1}\sum_{\beta=2}^\mathrm{N} \tilde{\pmv}_\beta^{\ce{m}} \cdot \Vflux_\beta
	= \charge \sum_{\beta=2}^{\ce{N}} \tilde{\pmv}_\beta^{\ce{m}} \cdot \Mflux_\beta,
	\\
	\label{eq:ConvertToMassReduc}
    \Mflux_\alpha - \Vflux_\alpha
	= \conc_\alpha \frac{M_1}{\density \pmv_1} \sum_{\beta=2}^\mathrm{N} \tilde{\pmv}_\beta^{\ce{m}} \cdot  \Vflux_\beta
	= \conc_\alpha \sum_{\beta=2}^\mathrm{N}\tilde{\pmv}_\beta^{\ce{m}} \cdot \Mflux_\beta.
\end{gather}
The relations \cref{eq:ConvertToMassReduc} hold for all flux densities $\Vflux_1,\ldots,\Vflux_{\ce{N}}$. Apparently, the electric flux densities are invariant under frame transformations in the electroneutral case ($\charge=0$).\cite{Miller1966a}

\subsection{Transference Numbers}
\label{sec:transference_number}

In this section, we focus on the concept of the transference numbers. We discuss different definitions for these transport coefficients, and derive the corresponding number of independent quantities. Here, we restrain our discussion mainly to the volume-based description. However, this discussion generalizes to any internal reference frame. Finally, we state the transformation rules for the transference numbers between different internal reference frames.

The N-1 transference numbers $t_\alpha^{\mathrm{v;red}}$ for an N-component electrolyte introduced above, see \cref{eq:volfluxreduced}, follow directly from the Onsager coefficients via $t_\alpha^{\mathrm{v;red}} = F^2\tilde{z}_\alpha/\kappa^{\mathrm{v}} \cdot \sum_{\beta=2}^{\ce{N}}\onsager_{\alpha\beta}^{\mathrm{v}} \tilde{z}_\beta^{\mathrm{v}}$ (see also \cref{eq:onsager}).\cite{schammer2020theory} By construction, they sum up to unity, $\sum_{\alpha=2}^{\mathrm{N}}t_\alpha^{\mathrm{v;red}}=1$, such that only N-2 independent parameters exist.

Commonly, transference numbers are understood as a measure for the ratio of current density that is carried by one species, \emph{i.e.} the contribution of $\Vflux_\alpha$ to the overall current density $\Jflux$. However, there are two important aspects regarding this interpretation. First, we emphasize that the flux densities $\Vflux_\alpha$ relate to the internal reference frame given by the volume-averaged drift velocity, \emph{i.e.} comprise a drift-correction $\Vflux_\alpha=\conc_\alpha\vel_\alpha-\conc_\alpha \vvel$ (in the electroneutral case, or if the drift velocity vanishes, $\Jflux^\uppsi=\sum_{\alpha=1}^{\ce{N}} Fz_\alpha \conc_\alpha \vel_\alpha - \charge\vel^\uppsi$ is the same in all frames, and equals the electric current defined relative to external coordinates,\cite{Miller1966a} \emph{i.e.} the lab-frame). Second, due to the relation \cref{eq:volfluxreduced}, the above interpretation of the transference numbers applies more directly to the quantity $t_\alpha^\mathrm{v,red}/ \tilde{z}_\alpha^\mathrm{v}$ than to the quantities $t_\alpha^\mathrm{v,red}$ itself. Also, the quantities $\tilde{z}_\alpha^{\mathrm{v}}$ can take counterintuitive values, as even uncharged species can get $ \tilde{z}_\alpha^{\mathrm{v}} \neq 0$.\cite{schammer2020theory} Hence, we argue that the N-1 flux ratios defined by 
\begin{equation}
	\tau_\alpha^\mathrm{v} = \frac{t_\alpha^\mathrm{v,red}}{\tilde{z}_\alpha^\mathrm{v}} = \frac{t_\alpha^\mathrm{v}}{z_\alpha}, \qquad \alpha \geq 2,
\end{equation}
are a more intuitive definition, which relates better to the physical interpretation from above. Beneath the N-1 flux ratios $\tau_\alpha^{\mathrm{v}}$, we introduced N-1 quantities $t_\alpha^{\mathrm{v}}$ for charged ion species, which are weighted by their non-zero valence.

In contrast to the N-1 transference numbers $t_\alpha^{\mathrm{v,red}}$, we can extend the set of N-1 quantities $(\tau_2,\ldots,\tau_{\ce{N}})$ and N-1 quantities $(t_2,\ldots,t_{\ce{N}})$ by a quantity $\tau_1$ and $t_1^{\mathrm{v}}$, relating to the flux contribution of $\Vflux_1$. For this purpose, we use \cref{eq:volfluxconstr}, $\Vflux_1=-\sum_{\alpha=2}^{\ce{N}} \pmv_\alpha/\pmv_1\cdot \Vflux_\alpha$, and express the flux densities $\Vflux_2,\ldots,\Vflux_{\ce{N}}$ via $\Jflux^{\mathrm{v}}$ assuming chemical equilibrium,
\begin{equation}
	\label{eq:t1_tau1}
	\tau_1^\mathrm{v} = \frac{t_1^\mathrm{v}}{z_1} = - \sum_{\alpha=2}^\mathrm{N}\tau_\alpha^\mathrm{v}\cdot  \frac{\pmv_\alpha}{\pmv_1} .
\end{equation}
As consequence, in total, there exist N quantities $\tau_\alpha^{\mathrm{v}}$ and N quantities  $t_\alpha^{\mathrm{v}}$. However, by construction, not all of these are independent since they are constrained by charge continuity and by the volumetric flux constraint, \emph{viz.} 
\begin{gather} 
	\label{eq:tzconstr}
	\sum_{\alpha=1}^{\ce{N}} t_\alpha^{\mathrm{v}} = \sum_{\alpha=1}^\mathrm{N} z_\alpha \tau_\alpha^{\mathrm{v}} = 1 \shortintertext{and} 
	\label{eq:tnuconstr}
	\sum_{\alpha=1}^{\mathrm{N}} \pmv_\alpha \frac{t_\alpha^{\mathrm{v}}}{z_\alpha} =
	\sum_{\alpha=1}^{\mathrm{N}} \pmv_\alpha \tau_\alpha^{\mathrm{v}}= 0.
\end{gather}
Hence, similar to the quantities $t_\alpha^{\mathrm{v,red}}$, only N-2 quantities $\tau_\alpha^{\mathrm{v}}$, and only N-2 quantities $t_\alpha^{\mathrm{v}}$ are independent. However, in contrast to $t_\alpha^{\mathrm{v,red}}$, there exist also quantities $\tau_1^{\mathrm{v}}$ and $t_1^{\mathrm{v}}$.

It has to be noted though, that, similar to the quantities $t_\alpha^{\mathrm{v;red}}$, the quantities $\tau^{\mathrm{v}}_\alpha$ measure the species contribution to the overall current density $\Jflux^{\mathrm{v}}$ in the volume-based reference frame, which differs from the electric current measured in the resting laboratory frame.

This discussion for the volume-based description generalizes to any other description based on an arbitrary internal frame of reference. In particular, the normalization  \cref{eq:tzconstr} holds in any frame, whereas \cref{eq:tnuconstr} takes a slightly different form in different frames (as consequence of differing flux constraints \cref{eq:universal_flux_constraint}). For example, in the CM-based description, we find constraints
\begin{gather}
	\label{eq:binary_CM_tau_constraints}
	\sum_{\alpha=1}^{\ce{N}} t_\alpha^{\mathrm{m}} = \sum_{\alpha=1}^\mathrm{N} z_\alpha \tau_\alpha^{\mathrm{m}} = 1,
	\shortintertext{and} 
	\label{eq:binary_CM_tau_constraints2}
	\sum_{\alpha=1}^{\mathrm{N}} M_\alpha \frac{t_\alpha^{\mathrm{m}}}{z_\alpha} =
	\sum_{\alpha=1}^{\mathrm{N}} M_\alpha \tau_\alpha^{\mathrm{m}}= 0.
\end{gather}
Another example is the internal reference frame with respect to the first species (where $\vel^\solventlabel=\vel_1$ defines the drift velocity). Here, the universal flux constraint implies $\Nflux^\solventlabel_1=0$ such that this flux does not contribute to the current $\Jflux$. In accordance with this fact, $t^\solventlabel_1 =\tau^\solventlabel_1= 0$ and, thus, the remaining N-1 quantities sum up to unity, $\sum_{\mathrm{N}=2}^\alpha t^\solventlabel_\alpha = \sum_{\alpha=2}^\mathrm{N} z_\alpha \tau_\alpha^{\solventlabel} = 1$.

Our description rationalizes the role of the transference numbers in an electrolyte with only two ion constituents. For such electrolytes, sign and magnitude of the transference numbers become arbitrary, as they are completely determined by the frame-specific form of the two constraints discussed above.\cite{sundheim1956transference,matsunaga2007velocity,Sinistri1962} In the volume-based description, the transference numbers are completely determined by the partial molar volumes via \cref{eq:tzconstr,eq:tnuconstr},\cite{Sinistri1962}
\begin{equation}
	t_1^{\mathrm{v}}=(1+\pmv_1/\pmv_2)^{-1}, \quad \textnormal{and} \quad  t_2^{\mathrm{v}}=(1+\pmv_2/\pmv_1)^{-1}.
\end{equation} 
An analogous relation, which is based on the molar masses, holds in the CM-based description (see \cref{eq:binary_CM_tau_constraints,eq:binary_CM_tau_constraints2}), 
\begin{equation}
	t_1^\mathrm{m}=(1+M_1/M_2)^{-1}, \quad \textnormal{and} \quad t_2^\mathrm{m}=(1+M_2/M_1)^{-1}.
\end{equation}
This result reproduces "Sundheims Golden rule".\cite{sundheim1956transference,matsunaga2007velocity}

Finally, we state the transformation rules of the transference numbers between the volume-based description and the CM-based description for electroneutral systems where $\charge=0$ (see \cref{sec:SItrf_transferencenumbers} in the ESI$^{\dag}$ for details),
\begin{gather} \label{eq:tConvertToMassReduc}
	t_\alpha^\mathrm{m} - t_\alpha^\mathrm{v} = \conc_\alpha z_\alpha \frac{M_1}{\density \pmv_1} \sum_{\beta=2}^\mathrm{N} \frac{\tilde{\pmv}_\beta^\mathrm{m}}{z_\beta} \cdot  t_\beta^\mathrm{v}
	= \conc_\alpha z_\alpha \sum_{\beta=2}^\mathrm{N} \frac{\tilde{\pmv}_\beta^{\ce{m}}}{z_\beta} \cdot t_\beta^\mathrm{m}.
\end{gather}
Apparently, the sign of the transference numbers is not conserved under frame transformations, because the quantities $z_\alpha\tilde{\pmv}_\beta^{\ce{m}}/z_\beta$ can become negative.

The corresponding transformation into the species-based reference frame depends upon the relative species concentrations (see \cref*{sec:SI_DerivTransformRules_vs} in the ESI.$^{\dag}$)
\begin{equation}
    \label{eq:tConvertToSpeciesFrame}
    t_\alpha^\solventlabel = t_\alpha^\mathrm{v} - \frac{z_\alpha \conc_\alpha}{z_1 \conc_1} t_1^\mathrm{v} = t_\alpha^\mathrm{m} - \frac{z_\alpha \conc_\alpha}{z_1 \conc_1} t_1^\mathrm{m}
\end{equation}
Note that, by construction, $t_1^\solventlabel = 0$.

\section{Application to eNMR Measurements} \label{sec:appltoeNMR}

In the following, we apply our description to eNMR experiments. We structure this chapter as follows. First, in \cref{sec:exp_set_up}, we briefly sketch the experimental set-up and introduce species mobilities. The experimental details can be found in Ref.~\citenum{Lorenz2022}. Next, in \cref{sec:transport_params}, we present our theoretical description of eNMR experiments, and state the underlying assumptions. Furthermore, we show how mobilities from eNMR are related to transference numbers and highlight the reference frame dependence of those transport parameters. In \cref{sec:BinarySys}, we shortly discuss the measurements on pure ILs and validate our assumption of a vanishing volume flux focusing on the theoretical aspects. In \cref{sec:TernarySys}, we consider electrolytes with three ion constituents. Instead of discussing the mobilities,\cite{Lorenz2022} we here put emphasis on the transference numbers, which are important transport parameters for continuum models. Finally, in \cref{sec:CompRefFrames} we compare transference numbers in different reference frames.

Note, that another possible application of our theory is given in the ESI$^{\dag}$, namely the modeling of electrochemical systems, e.g. batteries, with porous electrode theory. We state the respective modeling equations in \cref*{sec:SI_porous}.

\subsection{Experimental Set-up and Species Mobilities}
\label{sec:exp_set_up}

In this section, we briefly sketch the experimental set-up of eNMR measurements and discuss the species mobilities. For more details regarding the experimental aspects, we refer to Refs.~\citenum{Lorenz2022}, \citenum{Gouverneur2015} and \citenum{Schmidt2020}.

In electrophoretic NMR, a canonical PFG NMR experiment is performed with additional electric field pulses. The external field $\efield^{\ce{ext}}$ induces a constant force upon the ions, such that after a negligibly short acceleration period a charged species will exhibit a constant drift velocity, which is proportional to the external field, $\vel_\alpha^{\ce{drift}}=\mobility_\alpha \efield^{\ce{ext}}$. The species mobilities $\mobility_\alpha$ can be obtained from varying the electric field strength $\efield^{\ce{ext}}$ and evaluating the phase shift of the NMR signal, which is caused by the uniform displacement of this species in space. As the encoding of space is performed by magnetic field gradients, the species velocities $\vel_\alpha$ are generally determined in the laboratory frame. We identify the drift velocity of the ions with the species velocities, such that $\mobility_\alpha \efield^{\ce{ext}} = \vel_\alpha^{\ce{drift}}=\vel_\alpha = \Vflux_\alpha/\conc_\alpha + \vvel$.

The electric conductivity $\kappa$ of all electrolyte mixtures was measured experimentally using impedance spectroscopy (see Ref. \citenum{Lorenz2022} for details) and the values are listed in \cref*{tab:SI_binarydata,tab:SI_ternarydata} in the ESI.$^{\dag}$

\subsection{Model: Assumptions and Parameters}
\label{sec:transport_params}
In this section, we state our model assumptions for this system and discuss the mobilities and  transport parameters.

We assume that, during the eNMR experiment, the bulk electrolyte remains electroneutral ($\charge = 0$) and that all concentration profiles are constant ($\nabla \chempot_\alpha^\mathrm{v} = 0$).\cite{Lorenz2022} Because this implies that the electric current density and the electric conductivity are equal in all frames, we omit the labels for these two quantities. Furthermore, we assume a boundary condition for the electric field of the electrolyte, where $\efield = \efield^{\ce{ext}}$. This constitutes chemical equilibrium, \emph{i.e.} $\Vflux_\alpha=t_\alpha^{\mathrm{v}}/F z_\alpha \cdot \Jflux$, where the electric current density simplifies to $\Jflux = -\kappa \nabla \elpot = \kappa \efield$. Altogether, we thus find a relation between the species mobilities and the transport parameters,
\begin{gather}
	\mobility_\alpha \efield = \frac{\kappa t_\alpha^{\mathrm{v}}}{F z_\alpha \conc_\alpha} \cdot \efield + \vvel, 
	\intertext{or, using $\vvel / \efield = \sum_{\beta=1}^{\ce{N}}\conc_\beta\pmv_\beta\mobility_\beta$,} 
	\label{eq:geenral_relation_mobility_trans}
	t^{\textnormal{v}}_\alpha  
	=  \frac{F \conc_\alpha z_\alpha}{\kappa}\left(
	\mobility_\alpha - \frac{\vvel}{\efield}
	\right)
	= \frac{F z_\alpha c_\alpha\mobility_\alpha}{\kappa}  \sum_{\beta\neq \alpha} \conc_\beta\pmv_\beta
	\left(1 - \frac{\mobility_\beta}{\mobility_\alpha}\right).    
\end{gather}
Note that a similar equation can be found in any other frame of reference specified by some drift velocity $\vel^{\uppsi}=\sum_{\beta=1}^{\ce{N}}\uppsi_\beta\vel_\beta$, \textit{viz.} $t^{\uppsi}_\alpha = F z_\alpha c_\alpha\mobility_\alpha \cdot \sum_{\beta\neq \alpha} \uppsi_\beta (1 - \mobility_\beta/\mobility_\alpha)/\kappa $. Below we will discuss the transference numbers of Li in the two other frames of reference described above. In particular, we find for the mass-based quantities 
 \begin{gather}
\label{eq:tmLi_from_mob}
    t_{\ce{Li}}^{\ce{m}} =  \frac{F \conc_{\ce{Li}} \mobility_{\ce{Li}}}{\kappa}  \left[ \frac{\density_{\ce{an}}}{\density}\left( 1 - \frac{\mobility_{\ce{an}}}{\mobility_{\ce{Li}}} \right) + \frac{\density_{\ce{cat}}}{\density} \left( 1 - \frac{\mobility_{\ce{cat}}}{\mobility_{\ce{Li}}} \right) \right],
    \intertext{and for the quantities based on the species frame of the common anions,}
\label{eq:vehicularanionframe}
    t_{\mathrm{Li}}^\mathrm{s} = \frac{F  \conc_{\ce{Li}} \mobility_{\ce{Li}} }{\kappa}\left( 1 - \frac{\mobility_{\ce{an}}}{\mobility_{\ce{Li}}} \right).
\end{gather}

Because electrolytes are hardly compressible,\cite{Dreyer2013} we model the electrolyte systems as incompressible. In this case, the drift velocity $\vvel$ is spatially homogeneous, see \cref{eq:conveqzero}.  As consequence, it is determined by the boundary conditions of the system.

The experimental set-up consists of long vertical glass capillaries for thermal convection control with electrodes below and above  immersed in a reservoir of electrolyte.\cite{Gouverneur2015} Because the measurement area of the eNMR measurement is located inside the capillaries, the electrodes itself are outside the expected range of influence. This suggests that we set the boundary conditions right on top and below the glass capillaries, thus excluding electrode effects. Because the excess liquid above the capillaries leads to strong hydrostatic pressures at the end of the capillaries, we expect that there is no volume flux out of the measurement area.

We define the transference numbers derived in the laboratory frame from eNMR and conductivity measurements as
\begin{equation}
	\label{eq:transportparams}
	t_\alpha^{\mathrm{eNMR}}
	= F\conc_\alpha z_\alpha\cdot \mobility_\alpha / \kappa.
\end{equation}
In the fixed laboratory frame, the flux densities are simply $\Vflux_\alpha = \conc_\alpha\vel_\alpha = n_\alpha$ and the current density is $\Jflux^{\mathrm{v}}=F\sum_{\alpha=1}^{\ce{N}}\conc_\alpha z_\alpha\vel_\alpha = \jlab$. Therefore, the quantities $\tau_\alpha^{\mathrm{eNMR}} = t_\alpha^{\mathrm{eNMR}}/z_\alpha$ measure the species contribution to the electric current density as observed in the laboratory frame, and the quantities $\tau^{\mathrm{eNMR}}_\alpha$ equal exactly the intuitive interpretation discussed in \cref{sec:transference_number}. In particular, no specific knowledge of the partial molar volumes is needed for the calculation of the $t_\alpha^{\mathrm{eNMR}}$.

We want to validate that the volume flux vanishes. Thus, we check the two constraints \cref{eq:tzconstr,eq:tnuconstr} on the transference numbers. With \cref{eq:transportparams} they transfer to, 
\begin{equation}
	\label{eq:MobVolConstr}
	\sum_{\alpha=1}^\mathrm{N} \pmv_\alpha \conc_\alpha \mobility_\alpha = 0
	,\quad \textnormal{and}\quad
	F \sum_{\alpha=1}^\mathrm{N} z_\alpha \conc_\alpha \mobility_\alpha = \kappa,
\end{equation}
in terms of species mobilities $\mobility_\alpha$. Hence, only N-2 mobilities are independent (this would be also true if $\vvel\neq 0$).

Apparently, the electric conductivity of the electrolyte is completely determined by the species mobilities and the species concentrations. Note that the experimental determination of the conductivity $\kappa$ is comparatively straightforward using, \emph{e.g.} impedance spectroscopy as in this work. Thus, the second  constraint in \cref{eq:MobVolConstr} can be used  conveniently for a consistency check of the results obtained from eNMR measurements, which is a standard procedure to ensure high quality eNMR measurements.\cite{Gouverneur2015,Gouverneur2018,Rosenwinkel2019} Furthermore, the second constraint in \cref{eq:MobVolConstr} implies that all transference numbers can be calculated from eNMR mobilities alone via (see \cref{eq:transportparams}) 
\begin{equation} 
    \label{eq:tvolfrommu1}
    t_\alpha^\mathrm{eNMR} 
    = \frac{z_\alpha \conc_\alpha \mobility_\alpha}{\sum_{\beta=1}^\mathrm{N} z_\beta \conc_\beta \mobility_\beta},
\end{equation}
The \cref{eq:transportparams,eq:tvolfrommu1} reproduce the expression used in the  experimental literature for the calculation of transference numbers from mobilities, see \emph{e.g.},  eq.~(7) in Ref.~\citenum{Gouverneur2018}.

\subsection{Validation: Pure Ionic Liquids}
\label{sec:BinarySys}
In this section, we apply our description to eNMR measurements of pure ionic liquids (ILs). First, we show that in our description of the binary case the two species mobilities are completely determined by the conductivity and either the partial molar volumes (according to the volume-based description) or by the conductivity and the molar masses (CM-based description). Second, we apply the volume- and the mass-based description to experimental data for different pure ILs. Finally, we discuss that the volume-based approach yields a better description for incompressible electrolytes.

We assume that the pure IL dissociates into two ionic species which are oppositely charged, $XY\to X^+  + Y^-$. We label the cation-species by $\conc_{X^+}=\conc_+$ and the anion-species by $\conc_{Y^-}=\conc_-$. In the electroneutral state, the two ion concentrations of a pure IL have the same bulk concentration $\conc^{\ce{b}}$ and valencies $z_+ = - z_-$. However, due to the Euler equation for the volume, \cref{eq:volconstr}, the bulk concentration is completely determined by the partial molar volumes, such that $\conc_+ = \conc_- = \conc^{\ce{b}}=1/(\pmv_++\pmv_-)$.

In our transport theory, the electric conductivity is the only independent transport parameter in an electrolyte with two ion constituents. In particular, the transference numbers become arbitrary in the case of pure ILs and can be determined by, \emph{e.g.} the molar masses (according to the mass-based description), or by the partial molar volumes (according to the volume-based description), see \cref{sec:transference_number}. In the volume description, the two constraints $\kappa = F \conc^{\ce{b}} z_+(\mobility_+ - \mobility_-)$ and $\pmv_+ \mobility_+ + \pmv_- \mobility_- = 0$ determine the species mobilities (see \cref{eq:MobVolConstr}), 
\begin{gather}
	\label{eq:BinaryPredictMobs}
	\mobility_+ = \pmv_-\cdot \kappa/F  \quad \text{and} \quad \mobility_- = - \pmv_+ \cdot  \kappa / F,
	\shortintertext{such that the absolute ratio of the mobilities is inverse to the ratio of the partial molar volumes}
	\label{eq:BinaryMobRel}
	\lvert\,\mobility_+ /\mobility_-\,\rvert = \pmv_-/\pmv_+.    
\end{gather}

A similar relation can be found in the mass-based description of a pure IL in electroneutral state, when we assume that $\mvel=0$. Here, the two constraints \label{eq:binary_CM_tau_constraints,eq:binary_CM_tau_constraints2} yield
\begin{equation}
	\label{eq:mass_binary_mobility}
	\lvert\,\mobility_+/\mobility_-\,\rvert = M_- / M_+.
\end{equation}
This makes the mobility ratio a simple and straightforward parameter to identify the relevant boundary condition.\cite{Lorenz2022,Schoenhoff2018}

\begin{figure}[tb]
	\centering
\includegraphics[width=0.45\textwidth]{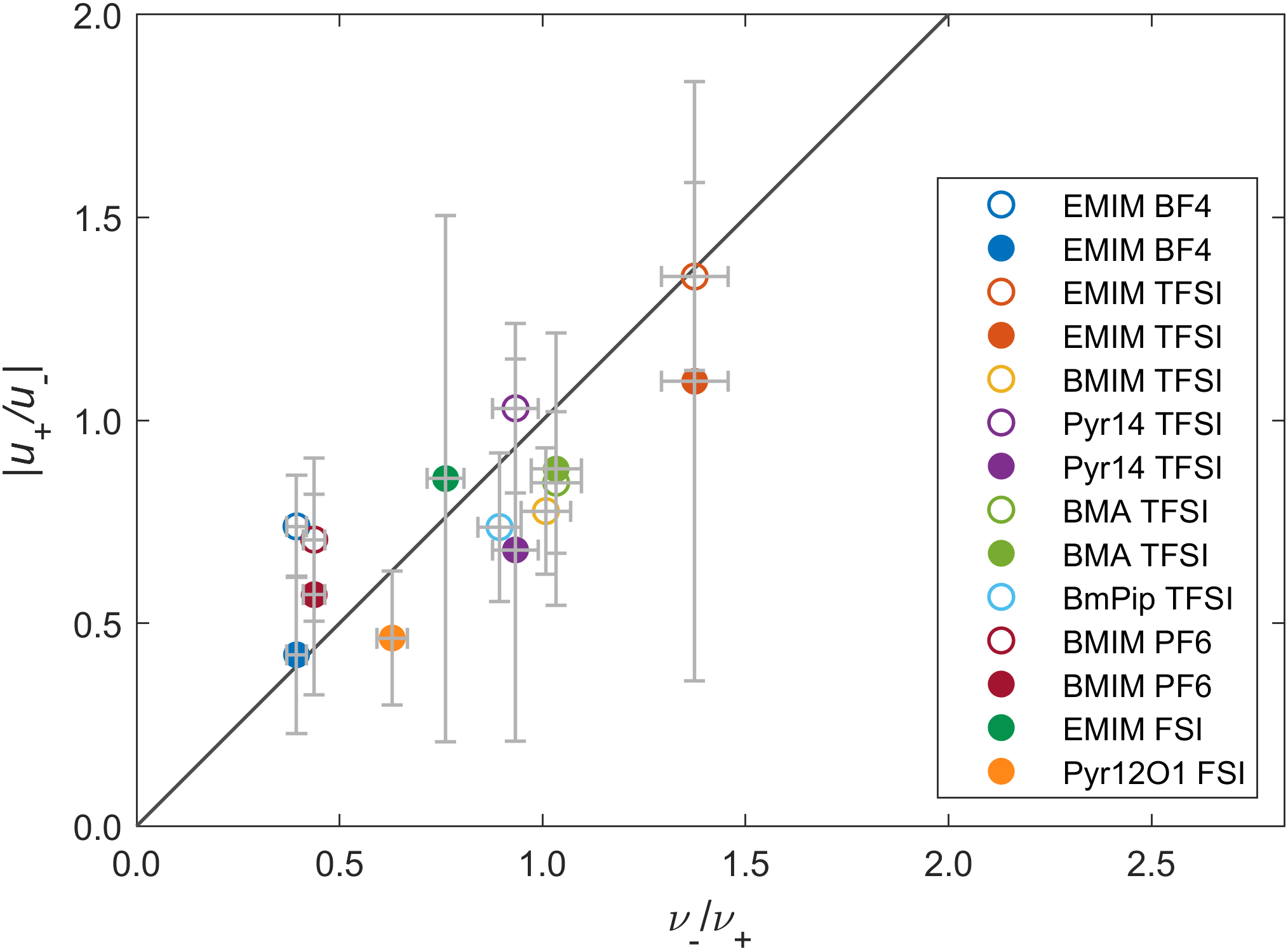}
	\caption{Ratio of mobilities versus ratio of partial molar volumes plotted for various ILs. Open circles: data from Gouverneur et al.\cite{Gouverneur2015}; Filled circles: data from Lorenz et al.\cite{Lorenz2022}. Diagonal line: analytical prediction (see  \cref{eq:BinaryMobRel}). The chemical acronyms are listed in \cref*{tab:SI_chem_acronyms} in the ESI.$^{\dag}$}
	\label{fig:BinaryMobsVsVols}
\end{figure}

We use these simple relations, \cref{eq:BinaryMobRel,eq:mass_binary_mobility}, to validate our assumption for the right boundary condition in comparison with experimental data for a wide range of different pure ILs. 

Here, we use the experimental data first published in Ref.~\citenum{Lorenz2022}, supplemented by data first published in Ref~\citenum{Gouverneur2015} (we restate relevant data in \cref*{tab:SI_binarydata}).$^{\dag}$ The partial molar volumes $\pmv_\alpha$ were consistently calculated from density measurements, as detailed in Ref.~\citenum{Lorenz2022}. The species mobilities $\mobility_\alpha$ were determined via eNMR measurements, see \cref{sec:exp_set_up}.

First, we probe the constraint \cref{eq:BinaryMobRel}. \Cref{fig:BinaryMobsVsVols} shows the absolute values for the mobility ratios $\lvert\mobility_+/\mobility_-\rvert$ as function of the relative partial molar volumes $\pmv_-/\pmv_+$ for the different IL electrolytes. The error bars stem from the systematic evaluation of all measurement uncertainties, see \cref*{sec:SI_pureILs}.$^{\dag}$ The diagonal line corresponds to the case where $\lvert\mobility_+/\mobility_-\rvert = \pmv_-/\pmv_+$. Apparently, for most systems, the results (including the error bars) lie on the diagonal line. This validates our assumption that $\vvel=0$.

In contrast, as we show in \cref*{fig:BinaryMobsVsMass} (see also Refs.~\citenum{Lorenz2022} and \citenum{Schoenhoff2018}), the mobility ratios do not align close to the diagonal line when plotted as functions of the ratio of molar masses. Hence, the experimental results  contradict the constraint in \cref{eq:mass_binary_mobility}. Thus, the assumption of vanishing drift velocity $\mvel$, \emph{i.e.} vanishing momentum flux, is badly chosen.

\begin{figure}[tb]
	\centering
	\includegraphics[width=0.45\textwidth]{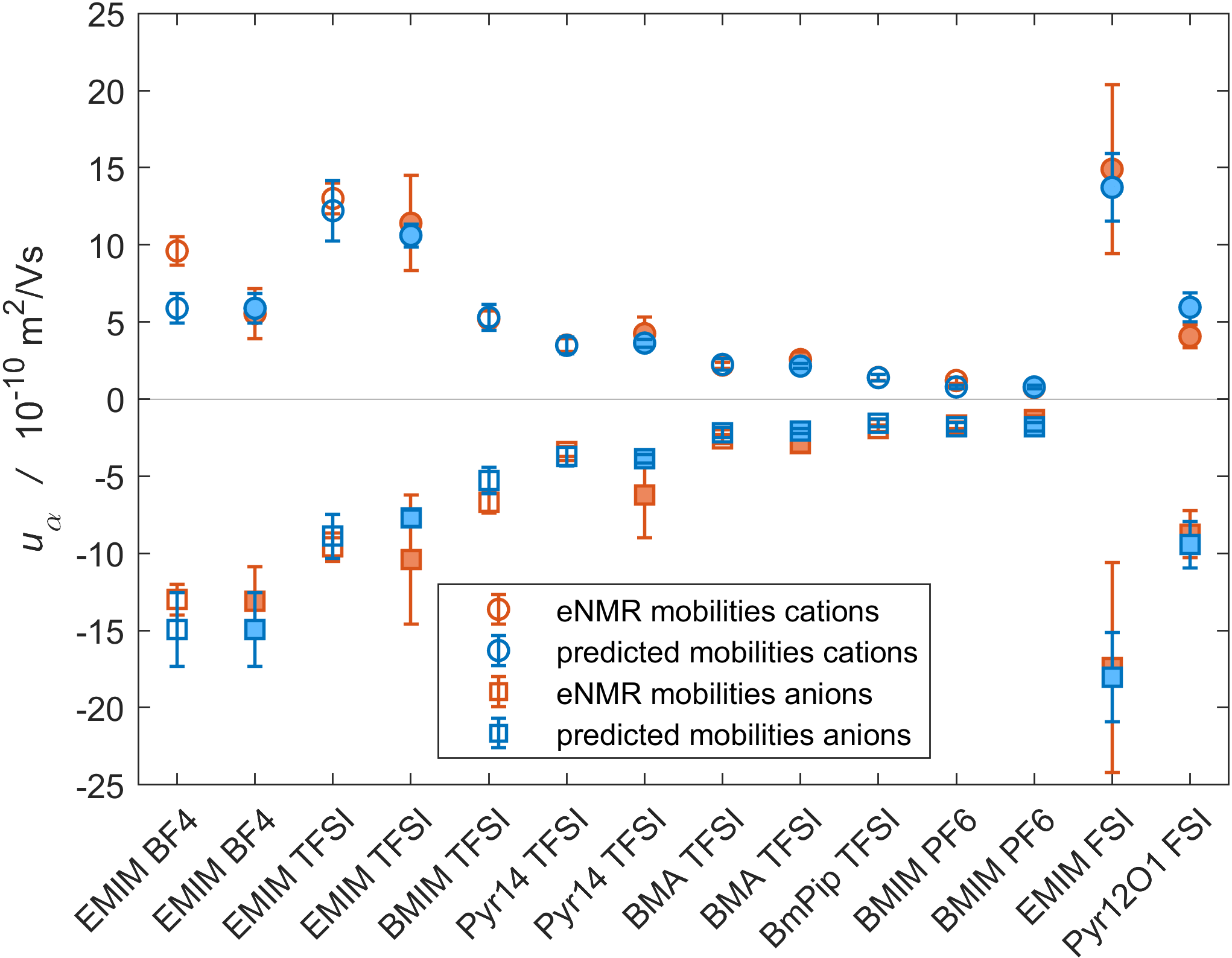}
	\caption{Comparison of the measured eNMR mobilities (orange) with the predicted mobilities from \cref{eq:BinaryPredictMobs} (blue) for various ILs; Circles: cation mobilities; Squares: anion mobilities; Open symbols: data from Gouverneur et al.\cite{Gouverneur2015}; Filled symbols: data from Lorenz et al.\cite{Lorenz2022} (c.f. \cref{fig:BinaryMobsVsVols}). The chemical acronyms are listed in \cref*{tab:SI_chem_acronyms} in the ESI.$^{\dag}$}
	\label{fig:BinaryMobseNMRvsPred}
\end{figure}

We highlight the relevance of the volume-based description for the eNMR experiment in \cref{fig:BinaryMobseNMRvsPred}, where we compare the analytical results for the species mobilities, as calculated from the partial molar volumes and the conductivity via \cref{eq:BinaryPredictMobs}, with the species mobilities obtained from eNMR measurements. \Cref{fig:BinaryMobseNMRvsPred} shows the corresponding set of eNMR mobilities for the ILs used in in \cref{fig:BinaryMobsVsVols}, where circles represent the positive cation values and squares the negative anion mobilities. Furthermore, experimental results are shown in orange and the analytically predicted mobilities in blue. Apparently, there is a high quantitative agreement between the experimental and theoretical results for the majority of the systems. This confirms our conclusion from above.

As summary, we find that the volume-based description subject to the assumption that $\vvel=0$ is quantitatively in very good agreement with the experimental data. In contrast, the mass-based description subject to the assumption of vanishing momentum flux does not reproduce the experimental results in a satisfactory manner. This confirms that the assumption of vanishing volume-based drift velocity is better justified than the assumption of local momentum conservation. As consequence, the volume-based model provides a better description for eNMR experiments in pure ILs than the mass-based description. Furthermore, we conclude that species diffusion in an incompressible electrolyte preserves, by definition, to a very good approximation local thermodynamic volume elements, but not the local momentum density. Indeed, in an incompressible mixture of components with different mass densities, volume conservation implies a center-of-mass motion and, thus, transport of momentum but no volume changes.
\\

\subsection{Validation: Ionic Liquid + Li-Salt Mixtures}
\label{sec:TernarySys}
Here, we focus on electrolyte mixtures composed of pure ILs with Li-salt with a common anion. In contrast to the previous section (where the transference numbers where arbitrary and we focused on the species mobilities), we now put emphasis on the transference numbers. We apply our volume-based description and validate our modeling assumptions concerning the relevant boundary condition for systems with three ionic constituents.

There exist three independent transport parameters in an electrolyte mixture composed of three different constituents. These are the electric conductivity, one transference number, and one diffusion coefficient. However, here we assume chemical equilibrium in the eNMR experiments, and thus neglect diffusion. As consequence, the set of independent transport parameters reduces to the electric conductivity $\kappa$ (which is frame-invariant in the electroneutral case, see \cref{sec:transport_params}), and one transference number (\emph{i.e.} $\tau^{\mathrm{v}}_3$ or $t_3^{\mathrm{v}}$ or $t_3^{\mathrm{v,red}}$) or one mobility.

For all electrolyte mixtures, we assume dissociation of the IL, $XY\to X^+ + Y^-$, and of the Li-salt $LiY\to Li^{+}+Y^-$. Thus, each electrolyte mixture gives rise to three independent ion-species. By convention, we label the common anion-species as first species by $\conc_{Y^-}=\conc_{\ce{an}}$, the cation-species of the IL as second species by $\conc_{X^+}=\conc_{\ce{cat}}$ and Li as third species by $\conc_{\ce{Li}}$. Hence, in our description, the only independent transference number is $t_{\ce{Li}}$.

In order to validate our model assumptions concerning a vanishing volume drift velocity for systems with three ion constituents, we check the two constraints \cref{eq:tzconstr,eq:tnuconstr} on the transference numbers obtained from eNMR via \cref{eq:transportparams} for each IL mixture stated in \cref*{tab:SI_ternarymixtures}.$^{\dag}$\cite{Gouverneur2018,Lorenz2022,Brinkkoetter2021} The $t^\mathrm{eNMR}_\alpha$ are listed in \cref*{tab:SI_ternarydata}.$^{\dag}$

\begin{figure}[tb]
	\centering
	\includegraphics[width=0.45\textwidth]{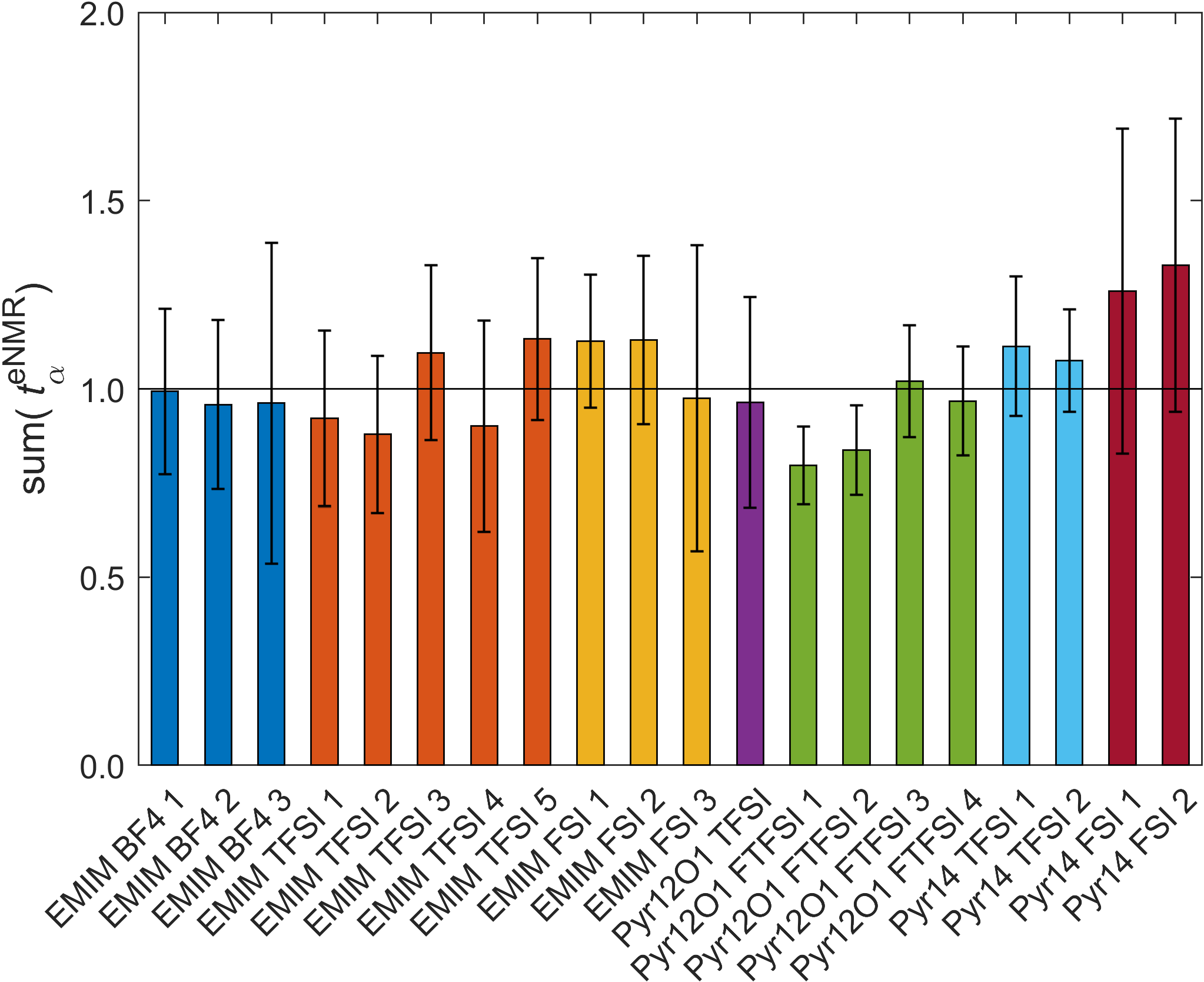}
	\caption{Probing charge conservation for various IL + Li-salt mixtures with common anion, see \cref{eq:tzconstr}. The horizontal line represents the analytical prediction. Data is based on mobilities from Refs.~\citenum{Gouverneur2018}, \citenum{Lorenz2022} and \citenum{Brinkkoetter2021}. Error bars result from measurement errors of the mobilities and conductivity. Chemical compositions are listed in \cref*{tab:SI_ternarymixtures} in the ESI and the $t^\mathrm{eNMR}_\alpha$ in \cref*{tab:SI_ternarydata}.$^{\dag}$}
	\label{fig:TernaryTsum1Dev}
\end{figure}
First, we focus on the constraint imposed by the assumption of charge conservation, $\sum_{\alpha=1}^{3} t_\alpha^{\mathrm{eNMR}} = 1$ (see \cref{eq:tzconstr}). \Cref{fig:TernaryTsum1Dev} illustrates this sum for all IL + Li-salt electrolyte mixtures listed in \cref*{tab:SI_ternarymixtures}.$^{\dag}$ The horizontal line marks the theoretically predicted value of one. To improve the readability, we group similar systems with varying salt concentrations by color. The assigned error bars stem mainly from errors made in eNMR measurements of the mobilities. Apparently, for nearly all systems the analytical prediction lies well within the error of the experimental results. However, for the Pyr$_\textnormal{12O1}$ FTFSI systems, a small deviation is clearly visible. We attribute this anomalous behaviour to an underestimation of the corresponding error made in the eNMR experiment. The partial molar volumes for the FTFSI$^-$ anion and, thus, also the density values had to be interpolated for those systems,\cite{Lorenz2022} resulting in slightly higher uncertainties than for the other mixtures. Apart from this singular deviation, there is in general a very good agreement with the analytical prediction. Note that the constraint \cref{eq:tzconstr} is already used as quality check for the eNMR measurements (see \cref{sec:transport_params} and Ref.~\citenum{Gouverneur2015,Gouverneur2018,Rosenwinkel2019}).

As second step, we probe the assumption of vanishing volume drift velocity ($\vvel=0$), which leads to the constraint in \cref{eq:tnuconstr} applying to the $t^\mathrm{eNMR}_\alpha$. For this purpose, we calculate the residual quantity
\begin{equation}
	\label{eq:ternarytransfvolconstr}
	\Delta 
    = \sum_{\alpha=1}^3 \frac{\pmv_\alpha}{\pmv_{\ce{total}}} \cdot \frac{t^\mathrm{eNMR}_\alpha}{z_\alpha}
	= \frac{F}{\kappa E} \cdot \frac{\vvel}{\pmv_{\ce{total}}}.
\end{equation}
This quantity measures the deviation from the analytical prediction, \textit{i.e.} $\Delta=0$. Here, $\pmv_{\ce{total}}=\sum_{\alpha=1}^{3}\pmv_\alpha$.

\begin{figure}[tb]
	\centering
	\includegraphics[width=0.45\textwidth]{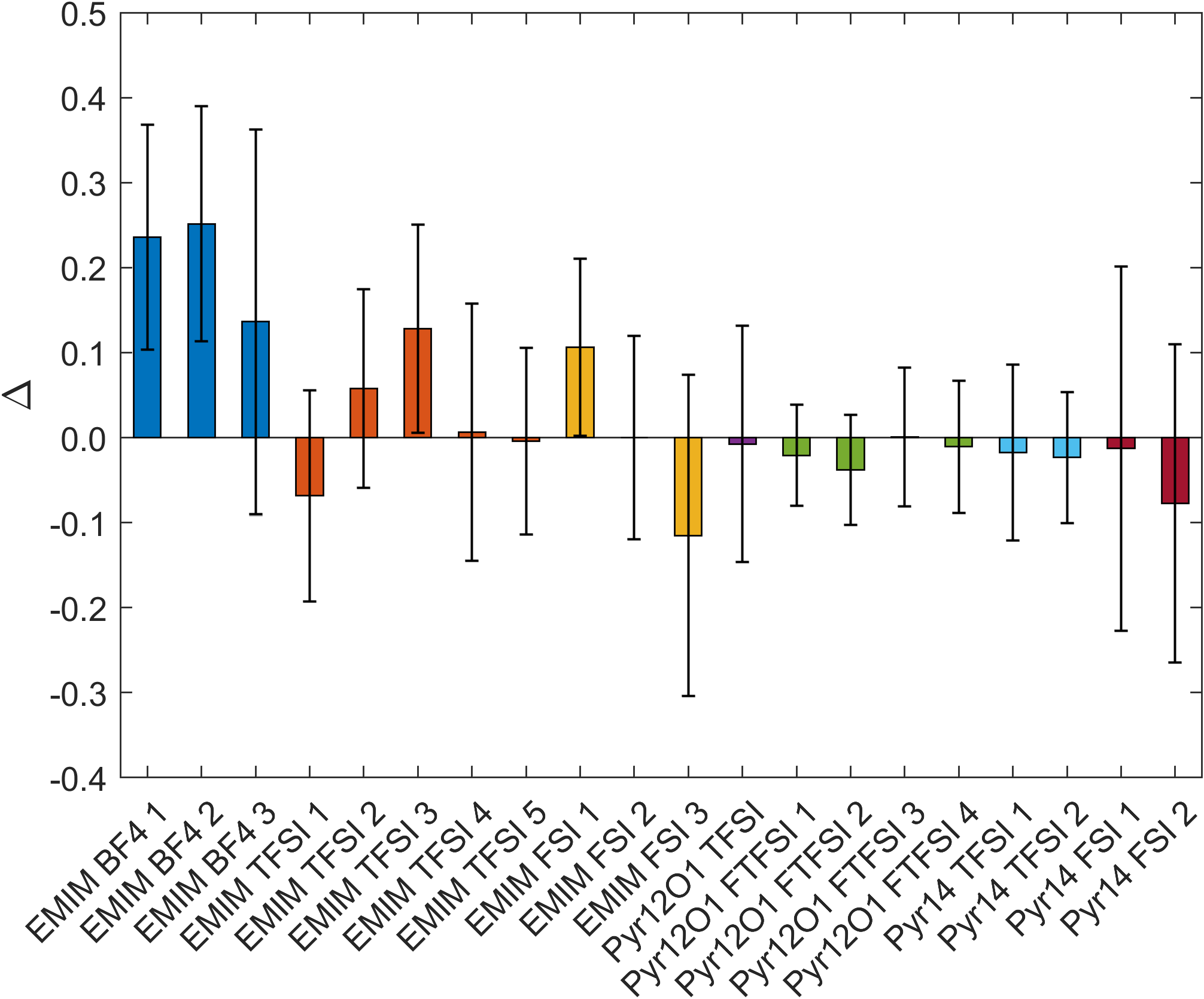}
	\caption{Probing the assumption of vanishing volume drift velocity for various IL + Li-salt mixtures with common anion, see \cref{eq:tnuconstr}. The horizontal line represents the analytical prediction. Data is based on mobilities from Refs.~\citenum{Brinkkoetter2018}, \citenum{Gouverneur2018}, \citenum{Lorenz2022} and \citenum{Brinkkoetter2021} with partial molar volumes from Ref.~\citenum{Lorenz2022}. Error bars result from measurement errors of the mobilities, conductivity and partial molar volumes. Chemical compositions are listed in \cref*{tab:SI_ternarymixtures} in the ESI and the $t^\mathrm{eNMR}_\alpha$ in \cref*{tab:SI_ternarydata}.$^{\dag}$}
	\label{fig:TernaryTnusum}
\end{figure}

\Cref{fig:TernaryTnusum} illustrates the quantity $\Delta$ for all IL + Li-salt electrolyte mixtures (see also \cref*{tab:SI_ternarymixtures}),$^{\dag}$ where similar systems with varying salt concentrations are grouped by color. Apparently, all up to two investigated systems fulfill the volume constraint very well. Only the first two EMIM BF$_4$ systems show a significant deviation from the analytical prediction (horizontal line at zero), which indicates a non-trivial drift velocity $\vvel$ for these systems. However, it was already argued in Ref. \citenum{Lorenz2022} that this deviation may result from side reactions occurring at the electrodes. If the partial molar volumes of the electrolyte species are different, these can cause unbalanced Faradaic productions of volume fractions, resulting in a volume drift velocity. Altogether, we conclude that the assumption of a vanishing volume-based drift velocity is justified. The $t^{\mathrm{eNMR}}_\alpha$ equal to a very good approximation the volume-based transference numbers $t^\mathrm{v}_\alpha$. Hence, \cref{eq:transportparams} is indeed correct, and the $t^{\mathrm{eNMR}}_\alpha$ can be interpreted as volume-based transference numbers with the boundary condition of $\vvel = 0$ in a standard eNMR set-up.

\subsection{Comparison of Reference Frames}
\label{sec:CompRefFrames}

\begin{figure}[tb]
	\centering
\includegraphics[width=0.45\textwidth]{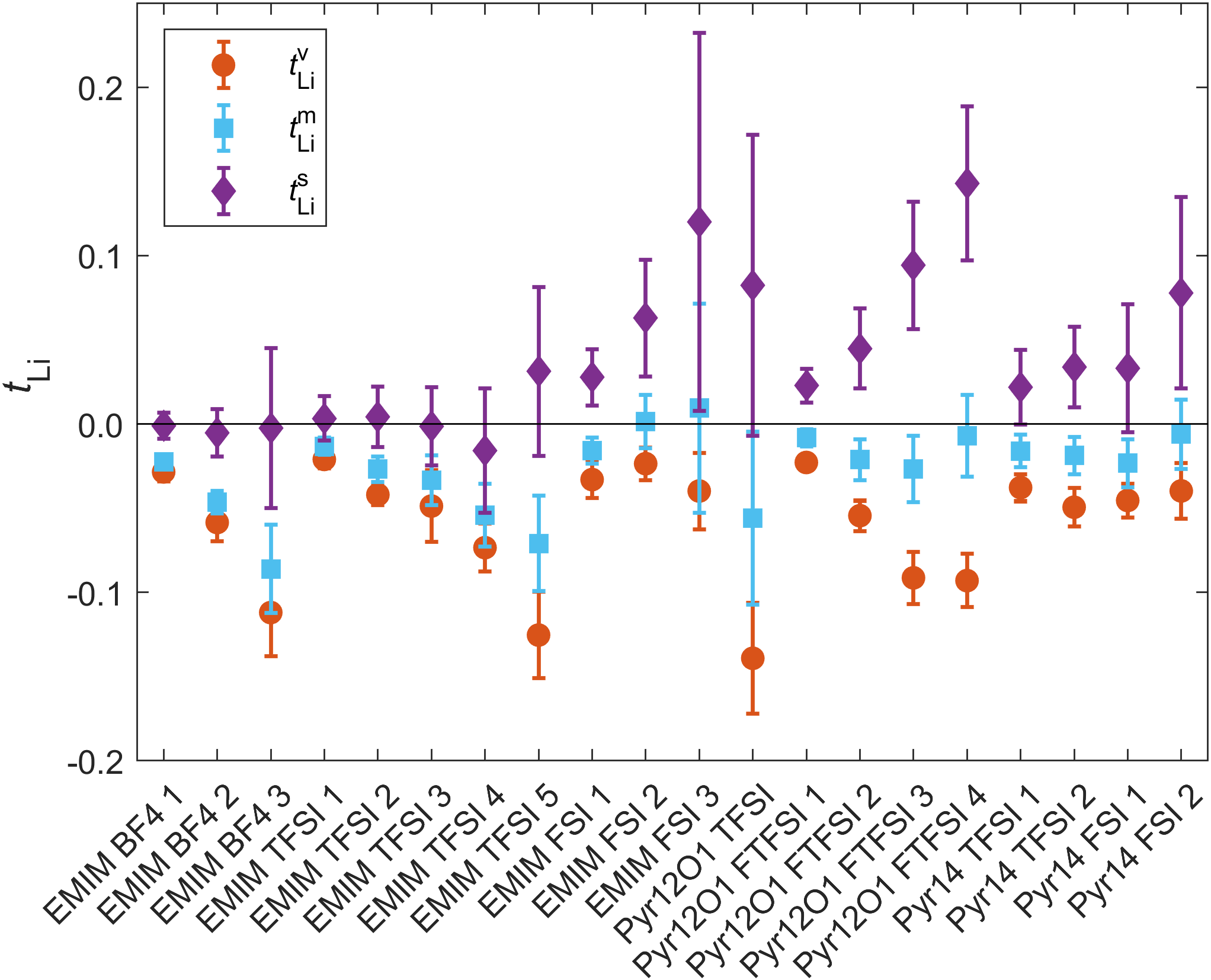}
	\caption{Comparison of sign and magnitude of Li transference numbers calculated with respect to the volume-based reference frame, \cref{eq:geenral_relation_mobility_trans}, the mass-based reference frame, \cref{eq:tmLi_from_mob}, and the anion-based reference frame, \cref{eq:vehicularanionframe}. Chemical compositions are listed in \cref*{tab:SI_ternarymixtures} in the ESI.$^{\dag}$}
	\label{fig:comp_mass_vol}
\end{figure}

In this section, we investigate the influence of the reference frame on the sign and magnitude of the Li transference numbers. For this purpose, in \cref{fig:comp_mass_vol}, we compare three different cases based on the volume description, \cref{eq:geenral_relation_mobility_trans}, the mass-based description, \cref{eq:tmLi_from_mob}, and the species-based description with respect to the common anion, \cref{eq:vehicularanionframe}.

First, we discuss the Li transference numbers in the volume-based description. The orange circles in \cref{fig:comp_mass_vol} show the transference numbers calculated from \cref{eq:geenral_relation_mobility_trans}. These quantities are negative for all electrolyte systems and their absolute magnitudes reside in the range between zero and roughly \num{0.15}.

Next, we discuss the transference numbers $t_{\ce{Li}}^{\ce{m}}$ of Li in the mass-based description, see the light blue squares in \cref{fig:comp_mass_vol}. We calculated the values from the mobilities via \cref{eq:tmLi_from_mob}. Apparently, the $t_\mathrm{Li}^{\mathrm{m}}$ tend to be more positive than the transference numbers in the volume-based description $t_\mathrm{Li}^{\mathrm{v}}$. However, an actual sign switch does only occur for two systems here (EMIM FSI 2 and EMIM FSI 3) and is, thus, not a dominant effect. Although for two more systems a possible switch of sign lies within the uncertainty when comparing the results of the volume- and the mass-based reference frame, the majority of transference numbers stays negative even within the error margins. \Cref{eq:tmLi_from_mob} shows that the relation of volume- and mass-based transference number depends on the ratios of anion and cation mobilities to Li mobilities as well as the ratio of anion to cation density (see the expression in square brackets).

Finally, we discuss the Li transference numbers $t_\mathrm{Li}^{\solventlabel}$ of the common anion reference frame shown by the purple diamonds in \cref{fig:comp_mass_vol}. The $t_\mathrm{Li}^{\solventlabel}$ are determined by the mobilities of Li and the common anions via \cref{eq:vehicularanionframe}. Most strikingly, the sign of the Li transference numbers for most of the investigated systems is positive, and only a few systems exhibit negative Li transference numbers in this reference frame. In addition, the absolute magnitudes spread over a relatively large range, when compared with the other two frames. We can rationalize the occurrence of positive Li transference numbers $t_{\ce{Li}}^{\ce{s}}$ using \cref{eq:vehicularanionframe}. Here, the mobilities $\mobility_{\ce{an}}$ and $\mobility_{\ce{Li}}$ are all negative, see Refs.~\citenum{Lorenz2022,Gouverneur2018,Brinkkoetter2021}. Thus, Li moves in the same direction as the anions, \emph{i.e.} the drift velocity. Furthermore, both of the two species, and hence the drift velocity, move into the opposite direction than the electric current. This seems to be in conflict with the transference numbers in \cref{fig:comp_mass_vol} being mostly positive (implying that Li moves in the same direction as the current). However, this seemingly contradictory behaviour can be resolved by the observation that the absolute magnitudes of the Li mobilities are in most cases smaller than the anion mobilities, see Refs. \citenum{Lorenz2022,Gouverneur2018,Brinkkoetter2021}, resulting in positive transference numbers $t_{\mathrm{Li}}^\mathrm{s} = F \conc_\mathrm{Li} \left( \lvert\mobility_{\ce{an}}\rvert - \lvert\mobility_{\ce{Li}}\rvert \right)/\kappa$. Hence, for such systems where the anions move "faster" than Li, Li moves into the same direction as the electric current when viewed from the co-moving anion-fixed frame, resulting in positive transference numbers $t_{\mathrm{Li}}^\solventlabel$. Indeed, for electrolyte mixtures where the $t_{\mathrm{Li}}^\solventlabel$ are negative, the experimental results show that the Li mobilities are larger in absolute magnitudes than the anion mobilities $\left( \lvert\mobility_{\ce{Li}}\rvert > \lvert\mobility_{\ce{an}}\rvert \right)$.

\section{Discussion}
\label{sec:discussion}
In this section, we complement our manuscript by a brief discussion of the relation between vehicular transport of Li and the sign of their transference numbers in the different frames of reference in IL + Li-salt electrolyte mixtures.

Our discussion relates to an ongoing debate in the literature, regarding the interpretation of negative transference numbers and the occurrence of highly correlated motion of Li-ions with oppositely charged ion-species. As argued by Schönhoff and co-workers,\cite{Brinkkoetter2018,Gouverneur2018,Brinkkoetter2021} vehicular transport of Li-ions with strongly positive Li-anion correlations implies that the anion and Li mobilities have the same sign (as the Li-ions are "travelling together" with the anions).

In our description, the transference numbers of Li with respect to any frame of reference are completely determined by the species mobilities, see \cref{eq:geenral_relation_mobility_trans,eq:tmLi_from_mob,eq:vehicularanionframe}. We make use of this description, and investigate the influence of negative mobilities $\mobility_{\ce{anion}}$ and $\mobility_{\ce{Li}}$ on the sign of the transference numbers in the three different frames of reference discussed above (the volume-based quantities $t_{\ce{Li}}^{\mathrm{v}}$, the mass-based quantities $t_{\ce{Li}}^{\mathrm{m}}$ and the anion-based quantities $t_{\ce{Li}}^{\mathrm{s}}$).

For all electrolyte systems studied in this work, the anion mobilities obtained from the eNMR measurements are negative. In addition, for all electrolyte systems, the mobilities of Li are negative as well, which indeed suggests a correlated motion of the Li-ions with the anions.

First, we discuss the influence of negative mobilities $\mobility_{\ce{an}}$ and $\mobility_{\ce{Li}}$ on the sign of $t_{\ce{Li}}^{\mathrm{v}}$. As we have shown in \cref{sec:TernarySys}, the $t_{\ce{Li}}^{\mathrm{eNMR}}$ calculated from \cref{eq:transportparams} equal to a very good approximation the $t_{\ce{Li}}^{\mathrm{v}}$ since the volume drift velocity vanishes, \emph{i.e.} $t_{\ce{Li}}^{\mathrm{eNMR}} \approx t_{\ce{Li}}^{\mathrm{v}}$. As a direct consequence of \cref{eq:transportparams}, vehicular transport implies that the sign of $t_{\ce{Li}}^{\mathrm{v}}$ equals the sign of $\mobility_{\ce{an}}$. Hence, vehicular transport yields negative Li transference numbers in the volume frame. This is in accordance with our results shown in \cref{fig:comp_mass_vol}.

Furthermore, in the common-anion frame, we have $\ce{sign}(t_{\ce{Li}}^{\solventlabel}) = \ce{sign}(\mobility_{\ce{Li}})\cdot (1- \lvert \mobility_{\ce{an}}/ \mobility_{\ce{Li}} \rvert)$ (see \cref{eq:vehicularanionframe}). Hence, depending on the relative magnitudes of $|\mobility_{\ce{Li}}|$ and $|\mobility_{\ce{an}}|$, vehicular transport either implies that $t_{\ce{Li}}^{\solventlabel}$ and $t_{\ce{Li}}^{\mathrm{v}}$ have opposite sign (if $\lvert\mobility_{\ce{an}}\rvert > \lvert\mobility_{\ce{Li}}\rvert)$, or the same sign (if $\lvert\mobility_{\ce{an}}\rvert < \lvert\mobility_{\ce{Li}}\rvert)$.
Here, in all systems, the mobilities of Li are negative, and the absolute magnitude of the anion mobility are mostly larger than that of Li. Hence, vehicular transport leads to positive transference numbers $t_{\ce{Li}}^{\solventlabel}$ in the common anion frame. This finding is in accordance with the results for the majority of systems shown in \cref{fig:comp_mass_vol}.

Finally, in the mass-based frame of reference it is harder to conclude from vehicular transport onto the sign of the Li transference numbers. As can be inferred from \cref{eq:tmLi_from_mob}, the relation between the sign of the Li transference number $t^{\ce{m}}_{\ce{Li}}$ and the anion mobility depends on the species densities and on the mobilities of the anions and Li.

\section{Conclusion}
\label{sec:Conclusion}

In this work, we supplement the discussion shown in Ref.~\citenum{Lorenz2022} and present a detailed discussion of transference numbers based on our novel transport theory. We applied our description to experimental results obtained from eNMR measurements and validated our model assumptions concerning the relevant boundary condition. 

Our discussion highlights the relevance of the reference frame for the interpretation of transport parameters. Furthermore, we present different definitions for the concept of transference numbers, and use universal constraints to identify the number of independent parameters. In addition to the external frame of reference given by the resting laboratory, we focus on three different internal reference frames defined by the center-of-mass motion, by the volume-based velocity, and by the species-based reference frame. We derive the exact transformation rules for the flux densities and transport parameters between these reference frames.

We apply our description of eNMR measurement devices to pure ILs and to electrolytes based on IL-Li-salt mixtures with common anions, as discussed in Ref.~\citenum{Lorenz2022}. Our analysis shows that, to a good approximation, the volume-based drift velocity can be neglected for such hardly compressible electrolytes. This highlights that for eNMR measurements the laboratory frame is best represented by the volume-based description. In contrast, assuming a vanishing momentum flux is a bad approximation.

It has to be noted though, that it can be advantageous to state transference numbers in the mass-based description. In particular, it is straightforward to obtain the Li transference numbers from the species mobilities of the eNMR measurements because the molar masses are the only additional parameters needed, see \cref{eq:tmLi_from_mob}. In contrast to the partial molar volumes (which are usually not so easy to determine as they may vary with composition, pressure and temperature) needed for the volume-based description, see \cref{eq:geenral_relation_mobility_trans}, the molar masses are unambiguously defined.

Furthermore, we show that transforming from volume- to mass-based description does not commonly yield a switch of sign of the transference numbers, although in rare cases this can occur. This picture is different when transforming into the common-anion reference frame. Here, most Li transference numbers are positive due to the high mobility of the anion.

Our description rationalizes an ongoing debate in the literature regarding sign and magnitude of transference numbers in ILs and highly concentrated electrolytes. Some apparent inconsistencies can be solved to a large extent when the boundary conditions are clearly stated and the choice of reference is clearly defined.


\nomenclature[I,01]{$+,\mathrm{cat}$}{Subscripts referring to cation}
\nomenclature[I,02]{$-,\mathrm{an}$}{Subscripts referring to anion}
\nomenclature[I,03]{$\mathrm{Li}$}{Subscript referring to Li}
\nomenclature[I,04]{$\alpha, \beta, \ldots$}{Greek subscripts refer to ion constituents}
\nomenclature[I,05]{$\uppsi$}{Superscript referring to $\uppsi$-based reference frame}
\nomenclature[I,06]{$\mathrm{m}$}{Superscript referring to mass-based reference frame}
\nomenclature[I,07]{$\solventlabel$}{Superscript referring to species-based reference frame}
\nomenclature[I,08]{$\mathrm{v}$}{Superscript referring to volume-based reference frame}

\nomenclature[P,01]{$D_{\alpha\beta}$}{Diffusion coefficient (with respect to thermodynamic driving force) \nomunit{\meter \squared \per \second}}
\nomenclature[P,02]{$F$}{Faraday constant \nomunit{\ampere \second \per \mol}}
\nomenclature[P,03]{$\onsager_{\alpha \beta}$}{Onsager matrix \nomunit{\second \mol \squared \per \kilogram \per \meter \cubed}}
\nomenclature[P,04]{$M_\alpha$}{Molar mass \nomunit{\kilogram \per \mol}}
\nomenclature[P,05]{N}{Number of species}
\nomenclature[P,06]{$t_\alpha$}{Transference number}
\nomenclature[P,07]{$\mobility_\alpha$}{Mobility \nomunit{\ampere \per \second \squared \per \kilogram}}
\nomenclature[P,08]{$z_\alpha$}{Valence / Charge number}
\nomenclature[P,09]{$\upbeta$}{Bruggemann coefficient}
\nomenclature[P,10]{$\epsilon$}{Porosity / Volume fraction of pore space}
\nomenclature[P,11]{$\upvarepsilon_{\ce{R}}$}{Relative permittivity}
\nomenclature[P,12]{$\upvarepsilon_0$}{Vacuum permittivity \nomunit{\ampere \second \per \volt \per \meter}}
\nomenclature[P,13]{$\kappa$}{Conductivity \nomunit{\ampere \squared \second \cubed \per \kilogram \per \meter \cubed}}
\nomenclature[P,14]{$\pmv_\alpha$}{Partial molar volume \nomunit{\meter \cubed \per \mol}}
\nomenclature[P,15]{$\tau_\alpha$}{Flux ratio / Transference number weighted by charge number}

\nomenclature[V,01]{$\conc_\alpha$}{Species concentration / Molarity \nomunit{\mol\per\meter\cubed}}
\nomenclature[V,02]{$\efield$}{Electric field \nomunit{\kilogram \meter \per \ampere \per \second \cubed}}
\nomenclature[V,03]{$I$}{Electric current \nomunit{\ampere}}
\nomenclature[V,04]{$\Jflux$}{Electric current density \nomunit{\ampere \per \meter \squared}}
\nomenclature[V,05]{$\jlab$}{Electric current density in the laboratory frame \nomunit{\ampere \per \meter \squared}}
\nomenclature[V,06]{$\Nflux_\alpha$}{Species flux density \nomunit{\mol \per \meter \squared \per \second}}
\nomenclature[V,07]{$n_\alpha$}{Species flux density in the laboratory frame \nomunit{\mol \per \meter \squared \per \second}}
\nomenclature[V,08]{$p$}{Pressure \nomunit{\kilogram \per \meter \per \second \squared}}
\nomenclature[V,09]{$\charge$}{Electric charge density \nomunit{\ampere \second \per \meter \cubed}}
\nomenclature[V,10]{$\entropy$}{Entropy production rate \nomunit{\kilogram \per \meter \per \second \cubed}}
\nomenclature[V,11]{$r_\alpha$}{species reaction rate \nomunit{\mol \per \meter \squared \per \second}}
\nomenclature[V,12]{$\vel$}{Drift velocity \nomunit{\meter \per \second}}
\nomenclature[V,13]{$\chempot_\alpha$}{Chemical potential \nomunit{\kilogram \meter \squared \per \second \squared \per \mol}}
\nomenclature[V,14]{$\density$}{Mass density \nomunit{\kilogram \per \meter \cubed}}
\nomenclature[V,15]{$\elpot$}{Electric potential \nomunit{\kilogram \meter \squared \per \ampere \per \second \cubed}}
\nomenclature[V,16]{$\varphi_\alpha$}{Electrochemical potential \nomunit{\kilogram \meter \squared \per \ampere \per \second \cubed}}

\printnomenclature

\section*{Conflicts of interest}
There are no conflicts to declare.

\section*{Acknowledgements}
This work was supported  by the European Union's Horizon 2020 research and innovation program via the "Si-DRIVE" project (grant agreement No 814464).



\balance

\renewcommand\refname{Notes and references}

\bibliography{bibliography} 
\bibliographystyle{rsc} 

\end{document}